\documentclass[fleqn]{jaes}

\usepackage{amsmath}
\usepackage{bm}
\usepackage{hyperref}

\usepackage{cite}
\usepackage{subfig}
\usepackage{amssymb}

\DeclareMathOperator*{\minimize}{minimize}

\begin{document}


\title{Weighted Pressure and Mode Matching\\for Sound Field Reproduction:\\Theoretical and Experimental Comparisons\thanks{To whom correspondence should be addressed, e-mail: koyama.shoichi@ieee.org.}}

\authorgroup{
\author{SHOICHI KOYAMA,\textsuperscript{1}}
\role{AES Member,}
\author{KEISUKE KIMURA,\textsuperscript{1}}
AND \author{NATSUKI UENO\textsuperscript{2}}
\email{(koyama.shoichi@ieee.org)\quad\quad\quad\quad\quad\quad\quad\quad\quad\quad\quad\quad\quad\quad\quad\quad\quad\quad\quad\quad\quad\quad\quad\quad (natsuki.ueno@ieee.org)}
\affil{\textsuperscript{1}Graduate School of Information Science and Technology, The University of Tokyo, Tokyo, Japan \\
\textsuperscript{2}Faculty of System Design, Tokyo Metropolitan University, Tokyo, Japan}
}

\abstract{%
Two sound field reproduction methods, weighted pressure matching and weighted mode matching, are theoretically and experimentally compared. The weighted pressure and mode matching are a generalization of conventional pressure and mode matching, respectively. Both methods are derived by introducing a weighting matrix in the pressure and mode matching. The weighting matrix in the weighted pressure matching is defined on the basis of the kernel interpolation of the sound field from pressure at a discrete set of control points. In the weighted mode matching, the weighting matrix is defined by a regional integration of spherical wavefunctions. It is theoretically shown that the weighted pressure matching is a special case of the weighted mode matching by infinite-dimensional harmonic analysis for estimating expansion coefficients from pressure observations. The difference between the two methods are discussed through experiments.
}
\maketitle


\section{INTRODUCTION}

The aim of sound field reproduction is to synthesize spatial sound using multiple loudspeakers (or secondary sources), which has various applications such as virtual/augmented reality audio, generation of multiple sound zones for personal audio, and noise cancellation in a spatial region. In some applications, the desired sound field to be reproduced is estimated using multiple microphones, which is called sound field capturing or estimation.

There are two major categories of sound field reproduction methods. One category includes analytical methods based on the boundary integral representations derived from the Helmholtz equation, such as \textit{wave field synthesis} and \textit{higher-order ambisonics}~\cite{Berkhout:JASA_J_1993,Spors:AES124conv,Poletti:J_AES_2005,Ahrens:Acustica2008,Wu:IEEE_J_ASLP2009,Koyama:IEEE_J_ASLP2013,Koyama:IEICE_J_EA_2014}. The other category includes numerical methods based on the minimization of a certain cost function defined for synthesized and desired sound fields inside a target region, such as \textit{pressure matching} and \textit{mode matching}~\cite{Kirkeby:JASA_J_1993,Nelson:J_SV_1993,Daniel:AES114conv,Poletti:J_AES_2005,Betlehem:JASA_J_2005,Ueno:IEEE_ACM_J_ASLP2019,Koyama:IEEE_ACM_J_ASLP2020}. Many analytical methods require the array geometry of loudspeakers to have a simple shape, such as a sphere, plane, circle, or line, and driving signals are obtained from a discrete approximation of an integral equation. In numerical methods, the loudspeaker placement can be arbitrary, and driving signals are generally derived as a closed-form least-squares solution. Pressure matching is based on synthesizing the desired pressure at a discrete set of control points placed over the target region. In mode matching, driving signals are derived so that the expansion coefficients of the spherical wavefunctions of the synthesized and desired sound fields are equivalent. Since the region in which the loudspeakers can be placed is limited in practical situations, a flexible loudspeaker array geometry in numerical methods will be preferable.

In this study, we theoretically and experimentally compare two numerical methods for sound field reproduction: \textit{weighted pressure matching}~\cite{Koyama:ICA2022} and \textit{weighted mode matching}~\cite{Ueno:IEEE_ACM_J_ASLP2019}. These two methods are derived by introducing a weighting matrix in the pressure and mode matching, respectively; therefore, they can be regarded as a generalization of the pressure and mode matching. The weighting matrix for the weighted pressure matching is derived on the basis of the kernel interpolation of the sound field~\cite{Ueno:IWAENC2018,Ueno:IEEE_J_SP_2021} from pressure at control points. In the weighted mode matching, the weighting matrix is defined as a regional integration of spherical wavefunctions. 
The relationship between pressure and mode matching has not been sufficiently elucidated from a theoretical perspective. We show that the weighted pressure matching is a special case of the weighted mode matching by combining with an infinite-dimensional harmonic analysis for sound field capturing~\cite{Ueno:IEEE_SPL2018,Ueno:IEEE_J_SP_2021}, starting with a common optimization problem.
Experimental evaluation comparing pressure/mode matching and weighted pressure/mode matching is carried out. The codes for reproducing the results are publicly available at \url{https://sh01k.github.io/MeshRIR/}. 

The rest of this paper is organized as follows. In Section~\ref{sec:prelim}, notations and basic theories on the sound field representation used throughout the paper are presented. The infinite-dimensional harmonic analysis for sound field capturing is also introduced. In Section~\ref{sec:prob}, the sound field reproduction problem is described. The weighted pressure and mode matching is formulated and theoretically compared in Section~\ref{sec:wpm-wmm}. Experimental comparisons are shown in Section~\ref{sec:exp}. In Section~\ref{sec:discuss}, differences between the two methods are discussed. Finally, Section~\ref{sec:conclusion} concludes this paper.


\section{NOTATIONS AND PRELIMINARIES}
\label{sec:prelim}

First, we provide several basic notations. Then, a sound field representation by spherical wavefunction expansion is introduced. We also briefly introduce a sound field capturing method based on infinite-dimensional harmonic analysis, which plays an important role in sound field reproduction methods. 

\subsection{Notations}

Italic letters denote scalars, lowercase boldface italic letters denote vectors, and uppercase boldface italic letters denote matrices. The sets of real and complex numbers are denoted by $\mathbb{R}$ and $\mathbb{C}$, respectively. Subscripts of scalars, vectors, and matrices indicate their indexes. To illustrate, the $(i,j)$th entry of the matrix $\bm{X}$ is represented as $x_{i,j}$. The imaginary unit and Napier's constant are denoted by $\mathrm{j}$ and $\mathrm{e}$, respectively. The complex conjugate, transpose, conjugate transpose, and inverse are denoted by superscripts $(\cdot)^{\ast}$, $(\cdot)^{\mathsf{T}}$, $(\cdot)^{\mathsf{H}}$, and $(\cdot)^{-1}$, respectively. The absolute value of a scalar $x$ and the Euclidean norm of a vector $\bm{x}$ are denoted by $|x|$ and $\|\bm{x}\|$, respectively. The absolute value for each element of matrix $\bm{X}$ is also denoted by $|\bm{X}|$.

The angular frequency, sound velocity, and wavenumber are denoted by $\omega$, $c$, and $k=\omega/c$, respectively. The harmonic time dependence $\mathrm{e}^{-\mathrm{j}\omega t}$ with the time $t$ is assumed according to conventions. 

\subsection{Expansion representation of sound field}

A solution of the homogeneous Helmholtz equation $u(\bm{r},\omega)$ of angular frequency $\omega$ at position $\bm{r}\in\mathbb{R}^3$ can be expanded around $\bm{r}_{\mathrm{o}}$ by using spherical wavefunctions~\cite{Williams:FourierAcoust,Martin:MultScat} as
\begin{align}
 u(\bm{r},\omega) &= \sum_{\nu=0}^{\infty} \sum_{\mu=-\nu}^{\nu} \mathring{u}_{\nu,\mu}(\bm{r}_{\mathrm{o}},\omega) \varphi_{\nu,\mu}(\bm{r}-\bm{r}_{\mathrm{o}},\omega) \notag\\
&= \bm{\varphi}(\bm{r}-\bm{r}_{\mathrm{o}},\omega)^{\mathsf{T}} \mathring{\bm{u}}(\bm{r}_{\mathrm{o}},\omega),
\label{eq:sphwave_exp}
\end{align}
where $\mathring{\bm{u}}(\bm{r}_{\mathrm{o}},\omega)\in\mathbb{C}^{\infty}$ and $\bm{\varphi}(\bm{r}-\bm{r}_{\mathrm{o}},\omega)\in\mathbb{C}^{\infty}$ are the infinite-dimensional vectors of expansion coefficients and spherical wavefunctions, respectively. The spherical wavefunction of the order $\nu$ and the degree $\mu$, $\varphi_{\nu,\mu}(\bm{r},\omega)$, is defined as
\begin{align}
 \varphi_{\nu,\mu}(\bm{r},\omega) = \sqrt{4\pi} j_{\nu}(k\|\bm{r}\|) Y_{\nu,\mu}\left(\frac{\bm{r}}{\|\bm{r}\|}\right),
\end{align}
where $j_{\nu}(\cdot)$ is the $\nu$th-order spherical Bessel function and $Y_{\nu,\mu}(\cdot)$ is the spherical harmonic function of order $\nu$ and degree $\mu$~\cite{Martin:MultScat}. The function $\varphi_{\nu,\mu}$ is scaled by the factor $\sqrt{4\pi}$ so that $\mathring{u}_{0,0}(\bm{r},\omega)$ corresponds to the pressure $u(\bm{r},\omega)$. Note that this scaling factor is not included in the standard definition of the spherical wavefunction. Hereafter, $\omega$ is omitted for notational simplicity. 

The translation operator $\bm{T}(\bm{r}_{\mathrm{o}}-\bm{r}_{\mathrm{o}}^{\prime}) \in \mathbb{C}^{\infty \times \infty}$ relates the expansion coefficients about two different expansion centers $\bm{r}_{\mathrm{o}}$ and $\bm{r}_{\mathrm{o}}^{\prime}$, i.e., $\mathring{\bm{u}}(\bm{r}_{\mathrm{o}})$ and $\mathring{\bm{u}}(\bm{r}_{\mathrm{o}}^{\prime})$, respectively, as~\cite{Martin:MultScat}
\begin{align}
 \mathring{\bm{u}}(\bm{r}_{\mathrm{o}}^{\prime}) = \bm{T}(\bm{r}_{\mathrm{o}}^{\prime}-\bm{r}_{\mathrm{o}}) \mathring{\bm{u}}(\bm{r}_{\mathrm{o}}),
\end{align}
where the element corresponding to the order $\nu$ and the degree $\mu$ of $\bm{T}(\bm{r})\mathring{\bm{u}}$, denoted as $[\bm{T}(\bm{r})\mathring{\bm{u}}]_{\nu,\mu}$, is defined as
\begin{align}
 &\left[ \bm{T}(\bm{r})\mathring{\bm{u}} \right]_{\nu,\mu} = \sum_{\nu^{\prime}=0}^{\infty} \sum_{\mu^{\prime}=-\nu^{\prime}}^{\nu^{\prime}} \left[ 
4\pi (-1)^{\mu^{\prime}} \mathrm{j}^{\nu-\nu^{\prime}} \right. \notag\\
&\left.
\cdot \sum_{l=0}^{\nu+\nu^{\prime}} \mathrm{j}^l j_l(k\|\bm{r}\|) Y_{l,\mu-\mu^{\prime}}\left(\frac{\bm{r}}{\|\bm{r}\|}\right) \mathcal{G}(\nu^{\prime},\mu^{\prime};\nu,-\mu,l)
\right] \mathring{u}_{\nu^{\prime},\mu^{\prime}}.
\end{align}
Here, $\mathcal{G}(\cdot)$ is the Gaunt coefficient. The translation operation is derived from the addition theorem of the spherical wavefunction~\cite{Martin:MultScat,Gumerov:FMM-HE}. The translation operator $\bm{T}(\bm{r}-\bm{r}^{\prime})$ has the following important properties:
\begin{align}
 &\bm{T}(-\bm{r}) =\bm{T}(\bm{r})^{-1} = \bm{T}(\bm{r})^{\mathsf{H}} \label{eq:trans_prop1}\\
 &\bm{T}(\bm{r}+\bm{r}^{\prime}) = \bm{T}(\bm{r}) \bm{T}(\bm{r}^{\prime}) \label{eq:trans_prop2}\\
 &\bm{\varphi}(\bm{r}-\bm{r}^{\prime})^{\mathsf{T}} \bm{T}(\bm{r}^{\prime}-\bm{r}^{\prime\prime}) = \bm{\varphi}(\bm{r}-\bm{r}^{\prime\prime})\label{eq:trans_prop3}.
\end{align}

\subsection{Sound field capturing based on infinite-dimensional harmonic analysis}
\label{sec:sf_est}

Here, we briefly introduce a method of estimating expansion coefficients of spherical wavefunctions of a sound field from microphone measurements~\cite{Ueno:IEEE_SPL2018}, i.e., sound field capturing/estimation method. Let $D \subseteq \mathbb{R}^3$ be a source-free target capturing region, and $M$ microphones are arbitrarily placed in $D$. The sound field capturing problem is to estimate the expansion coefficients at the position $\bm{r} \in D$, $\mathring{\bm{u}}(\bm{r})$, using the observed signal of the microphones $s_m$ at the positions $\bm{r}_{\mathrm{m},m} \in D$ ($m\in\{1,\ldots,M\}$). 

The microphone directivity patterns are assumed to be given as their expansion coefficients $c_{m,\nu,\mu}$ of spherical harmonic functions. By denoting the infinite-dimensional vector of the expansion coefficients $c_{m,\nu,\mu}$ by $\bm{c}_m\in\mathbb{C}^{\infty}$, we describe the observed signal $s_m$ as the inner product of $\bm{c}_{m}$ and $\mathring{\bm{u}}(\bm{r}_{\mathrm{m},m})$ as
\begin{align}
 s_m &= \sum_{\nu=0}^{\infty} \sum_{\mu=-\nu}^{\nu} c_{m,\nu,\mu}^{\ast} \mathring{u}_{\nu,\mu}(\bm{r}_{\mathrm{m},m}) \notag\\
&= \bm{c}_m^{\mathsf{H}} \mathring{\bm{u}}(\bm{r}_{\mathrm{m},m}) \notag\\
&= \bm{c}_m^{\mathsf{H}} \bm{T}(\bm{r}_{\mathrm{m},m}-\bm{r}) \mathring{\bm{u}}(\bm{r}),
\label{eq:sm_ip}
\end{align}
where the translation operator is used in the last line to relate $s_m$ with $\mathring{\bm{u}}(\bm{r})$. See Appendix for the derivation of the first line. Equation~\eqref{eq:sm_ip} can be rewritten as 
\begin{align}
 \bm{s} = \bm{\Xi}(\bm{r})^{\mathsf{H}} \mathring{\bm{u}}(\bm{r}),
\end{align}
where $\bm{s}=[s_1,\ldots,s_M]^{\mathsf{T}}\in\mathbb{C}^M$ and $\bm{\Xi}(\bm{r})\in\mathbb{C}^{\infty \times M}$ is described as
\begin{align}
 \bm{\Xi}(\bm{r}) &= 
\begin{bmatrix}
 (\bm{c}_1^{\mathsf{H}} \bm{T}(\bm{r}_{\mathrm{m},1}-\bm{r}))^{\mathsf{H}}, & \ldots, & (\bm{c}_M^{\mathsf{H}} \bm{T}(\bm{r}_{\mathrm{m},M}-\bm{r}))^{\mathsf{H}}
\end{bmatrix} \notag\\
&=
\begin{bmatrix}
 \bm{T}(\bm{r}-\bm{r}_{\mathrm{m},1})\bm{c}_1, & \ldots, & \bm{T}(\bm{r}-\bm{r}_{\mathrm{m},M})\bm{c}_M
\end{bmatrix}. 
\label{eq:est_xi}
\end{align}
Here, the property of the translation operator \eqref{eq:trans_prop1} is used. The expansion coefficient $\mathring{\bm{u}}(\bm{r})$ is estimated as
\begin{align}
 \mathring{\bm{u}}(\bm{r}) = \bm{\Xi}(\bm{r}) \left( \bm{\Psi} + \xi \bm{I} \right)^{-1} \bm{s},
\label{eq:est}
\end{align}
where $\xi$ is a constant parameter and $\bm{\Psi}:=\bm{\Xi}(\bm{r})^{\mathsf{H}}\bm{\Xi}(\bm{r})\in\mathbb{C}^{M \times M}$. From the property in Eq.~\eqref{eq:trans_prop2}, the $(m,m^{\prime})$th element of $\bm{\Psi}$ becomes
\begin{align}
 (\bm{\Psi})_{m,m^{\prime}} &= \bm{c}_m^{\mathsf{H}} \bm{T}(\bm{r}_{\mathrm{m},m}-\bm{r}) \bm{T}(\bm{r}-\bm{r}_{\mathrm{m},m^{\prime}}) \bm{c}_{m^{\prime}} \notag\\
&= \bm{c}_m^{\mathsf{H}} \bm{T}(\bm{r}_{\mathrm{m},m}-\bm{r}_{\mathrm{m},m^{\prime}}) \bm{c}_{m^{\prime}}.
\label{eq:est_psi}
\end{align}
Therefore, $\bm{\Psi}$ does not depend on the position $\bm{r}$ and depends only on the microphones' positions and directivities. Since the microphone directivity $c_{m,\nu,\mu}$ is typically modeled by low-order coefficients, Eq.~\eqref{eq:est_psi} can be simply computed in practice.  

Next, we consider estimating the pressure distribution $u(\bm{r})=\mathring{u}_{0,0}(\bm{r})$ using pressure microphones. The expansion coefficient of the directivity, $c_{m,\nu,\mu}$, is written as
\begin{align}
 c_{m,\nu,\mu} = 
\begin{cases}
 1, & \nu=0, \mu=0 \\
 0, & \text{otherwise}
\end{cases}.
\label{eq:c_omni}
\end{align}
Then, estimation Eq.~\eqref{eq:est} can be simplified as
\begin{align}
 u(\bm{r}) = \bm{\kappa}(\bm{r})^{\mathsf{T}} \left( \bm{K} + \xi \bm{I} \right)^{-1} \bm{s},
\end{align}
where 
\begin{align}
 \bm{K} &= 
\begin{bmatrix}
 j_0(k\|\bm{r}_1-\bm{r}_1\|) & \cdots & j_0(k\|\bm{r}_1-\bm{r}_M\|) \\
 \vdots & \ddots & \vdots \\
 j_0(k\|\bm{r}_M-\bm{r}_1\|) & \cdots & j_0(k\|\bm{r}_M-\bm{r}_M\|) 
\end{bmatrix}
\\
 \bm{\kappa}(\bm{r}) &= 
\begin{bmatrix}
 j_0(k\|\bm{r}-\bm{r}_1\|) & \ldots & j_0(k\|\bm{r}-\bm{r}_M\|) 
\end{bmatrix}^{\mathsf{T}}.
\end{align}
This equation can be regarded as kernel ridge regression with the kernel function of the 0th-order spherical Bessel function, which enables us to interpolate pressure distribution in a three-dimensional (3D) space with the constraint that $u(\bm{r})$ satisfies the Helmholtz equation~\cite{Ueno:IWAENC2018}. In a two-dimensional (2D) sound field, the kernel function is replaced with the 0th-order Bessel function.

In the sound field capturing, it is frequently impractical to capture the sound field in a large region using a single large microphone array, such as a spherical array. The estimation method described above enables us to use arbitrarily placed microphones, for example, distributed microphones~\cite{Iijima:JASA_J_2021}. Such a sound field capturing system will be useful in practical situations because of its flexibility and scalability. 

\section{SOUND FIELD REPRODUCTION PROBLEM}
\label{sec:prob}

\begin{figure}[t]
\centering
\includegraphics[width=55mm]{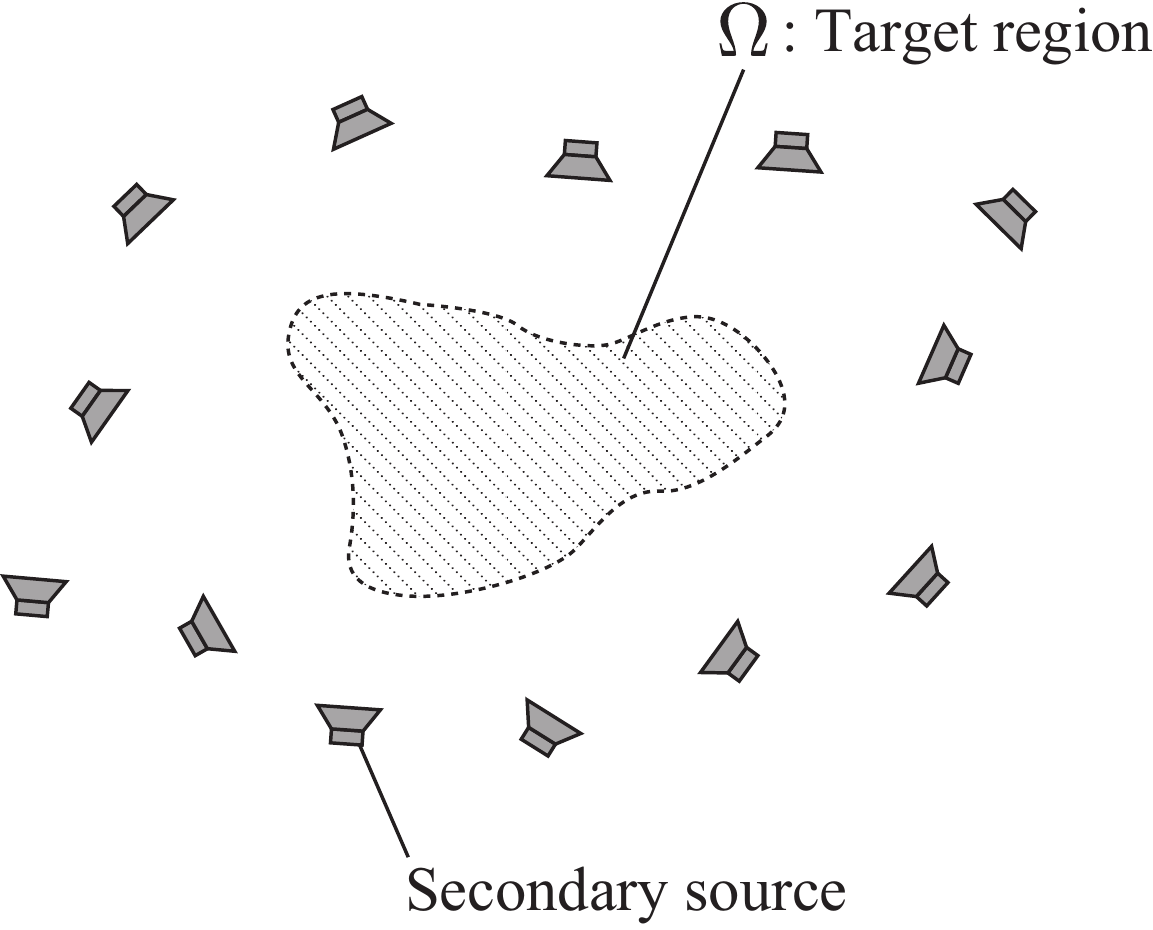}
\caption{The desired sound field is synthesized inside the target region $\Omega$ using multiple secondary sources.}
\label{fig:sfc}
\end{figure}

Suppose that $L$ secondary sources (loudspeakers) are placed around a target reproduction region $\Omega \subset \mathbb{R}^3$ as shown in Fig.~\ref{fig:sfc}. The desired sound field at $\bm{r}\in\Omega$ is denoted by $u_{\mathrm{des}}(\bm{r})$ in the frequency domain. The sound field $u_{\mathrm{syn}}(\bm{r})$ synthesized using the secondary sources is represented as
\begin{align}
 u_{\mathrm{syn}}(\bm{r}) = \sum_{l=1}^L d_l g_l(\bm{r}),
\end{align}
where $d_l$ is the driving signal of the $l$th secondary source, and $g_l(\bm{r})$ is the transfer function from the $l$th secondary source to the position $\bm{r}$ ($l\in\{1,\ldots,L\}$). The transfer functions $g_l(\bm{r})$ are assumed to be known by measuring or modeling them in advance. The goal of sound field reproduction is to obtain $d_l$ of the $L$ secondary sources so that $u_{\mathrm{syn}}(\bm{r})$ coincides with $u_{\mathrm{des}}(\bm{r})$ inside $\Omega$. 

We define the cost function to determine the driving signal $d_l$ for $l\in\{1,\ldots,L\}$ as
\begin{align}
 J &= \int_{\Omega} \left| \sum_{l=1}^L d_l g_l(\bm{r}) - u_{\mathrm{des}}(\bm{r}) \right|^2 \mathrm{d} \bm{r} \notag\\
&= \int_{\Omega} \left| \bm{g}(\bm{r})^{\mathsf{T}} \bm{d} - u_{\mathrm{des}}(\bm{r}) \right|^2 \mathrm{d}\bm{r},
\label{eq:cost}
\end{align}
where $\bm{g}(\bm{r})=[g_1(\bm{r}),\ldots,g_L(\bm{r})]^{\mathsf{T}}\in\mathbb{C}^L$ and $\bm{d}=[d_1,\ldots,d_L]^{\mathsf{T}}\in\mathbb{C}^L$ are the vectors of the transfer functions and driving signals, respectively. The optimal driving signal $\bm{d}$ can be obtained by solving the minimization problem of $J$. The cost function $J$ is formulated as the mean square error of the reproduction over the region $\Omega$. To incorporate the expected regional accuracy, a weighting function $\rho(\bm{r})$ ($\bm{r}\in\Omega$) is sometimes used as~\cite{Ueno:IEEE_ACM_J_ASLP2019}
\begin{align}
 J_{\rho} = \int_{\Omega} \rho(\bm{r}) \left| \bm{g}(\bm{r})^{\mathsf{T}} \bm{d} - u_{\mathrm{des}}(\bm{r}) \right|^2 \mathrm{d}\bm{r}.
\end{align}
The function $\rho(\bm{r})$ is designed on the basis of the regional importance of the reproduction accuracy. However, in this study, we focus on the case of a uniform distribution, i.e., $\rho(\bm{r})=1$, for simplicity.

\section{WEIGHTED PRESSURE AND MODE MATCHING}
\label{sec:wpm-wmm}

Several methods of approximately solving the minimization problem of Eq.~\eqref{eq:cost} have been proposed. We introduce two sound field reproduction methods, weighted pressure matching and weighted mode matching. 

\subsection{Weighted pressure matching}
\label{sec:wpm}

A simple strategy to solve the minimization problem of Eq.~\eqref{eq:cost} is to discretize the target region $\Omega$ into multiple control points, which is referred to as the pressure-matching method. Assume that $N$ control points are placed over $\Omega$ and their positions are denoted by $\bm{r}_{\mathrm{c},n}$ ($n\in\{1,\ldots,N\}$). The cost function $J$ is approximated as the error between the synthesized and desired pressures at the control points. The optimization problem of pressure matching is described as 
\begin{align}
 \minimize_{\bm{d}\in\mathbb{C}^L} \| \bm{Gd} - \bm{u}^{\mathrm{des}} \|^2 + \eta \|\bm{d}\|^2,
\label{eq:pm_prob}
\end{align}
where $\bm{u}^{\mathrm{des}} = [u_{\mathrm{des}}(\bm{r}_{\mathrm{c},1}), \ldots, u_{\mathrm{des}}(\bm{r}_{\mathrm{c},N})]^{\mathsf{T}} \in \mathbb{C}^N$ is the vector of the desired sound pressures and $\bm{G}=[\bm{g}(\bm{r}_{\mathrm{c},1}), \ldots, \bm{g}(\bm{r}_{\mathrm{c},N})]^{\mathsf{T}}\in\mathbb{C}^{N \times L}$ is the transfer function matrix between $L$ secondary sources and $N$ control points. The second term is the regularization term to prevent an excessively large amplitude of $\bm{d}$, and $\eta$ is a constant parameter. The solution of Eq.~\eqref{eq:pm_prob} is obtained as 
\begin{align}
 \bm{d}_{\mathrm{PM}} = \left( \bm{G}^{\mathsf{H}}\bm{G} + \eta\bm{I} \right)^{-1} \bm{G}^{\mathsf{H}} \bm{u}^{\mathrm{des}}.
\label{eq:pm_sol}
\end{align}

Owing to the discrete approximation, the cost function of pressure matching is formulated so that the synthesized pressure corresponds to the desired pressure only at the control points. Therefore, the region between the control points is not taken into consideration. 
When the distribution of the control points is sufficiently dense, the pressure values at the control points are sufficient to represent the sound field in the target region. However, since the pressures at the control points are measured by microphones in practice, small number of control points is preferable. Therefore, we consider approximating the cost function $J$ by interpolating the sound field from the pressures at the control points.
On the basis of the kernel interpolation introduced in Section~\ref{sec:sf_est}, $g_l(\bm{r})$ and $u_{\mathrm{des}}(\bm{r})$ are interpolated from those at the control points as
\begin{align}
 \hat{g}_l(\bm{r}) &= \bm{\kappa}_{\mathrm{c}}(\bm{r})^{\mathsf{T}} \left( \bm{K}_{\mathrm{c}} + \xi \bm{I} \right)^{-1} \bm{g}_{\mathrm{c},l} \\
 \hat{u}_{\mathrm{des}} (\bm{r}) &= \bm{\kappa}_{\mathrm{c}}(\bm{r})^{\mathsf{T}} \left( \bm{K}_{\mathrm{c}} + \xi \bm{I} \right)^{-1} \bm{u}^{\mathrm{des}},
\end{align}
where $\bm{g}_{\mathrm{c},l}$ ($\in\mathbb{C}^N$) is the $l$th column vector of $\bm{G}$, and $\bm{K}_{\mathrm{c}}\in\mathbb{C}^{N \times N}$ and $\bm{\kappa}_{\mathrm{c}}\in\mathbb{C}^N$ are respectively the matrix and vector consisting of the kernel function defined with the positions $\{\bm{r}_{\mathrm{c},n}\}_{n=1}^N$. Then, the cost function $J$ can be approximated as
\begin{align}
J &\approx \int_{\Omega} \left| \sum_{l=1}^L d_l \hat{g}_l(\bm{r}) - \hat{u}_{\mathrm{des}}(\bm{r}) \right|^2 \mathrm{d}\bm{r} \notag\\
&= \int_{\Omega} \left| \bm{\kappa}_{\mathrm{c}}(\bm{r})^{\mathsf{T}} \left( \bm{K}_{\mathrm{c}} + \xi \bm{I} \right)^{-1} \left(\bm{Gd} - \bm{u}^{\mathrm{des}} \right) \right|^2 \mathrm{d}\bm{r} \notag\\
&= \left(\bm{Gd} - \bm{u}^{\mathrm{des}} \right)^{\mathsf{H}} \bm{W}_{\mathrm{PM}} \left(\bm{Gd} - \bm{u}^{\mathrm{des}} \right),
\label{eq:cost_wpm}
\end{align}
where $\bm{W}_{\mathrm{PM}}$ is defined as
\begin{align}
 \bm{W}_{\mathrm{PM}} := \bm{P}^{\mathsf{H}} \int_{\Omega} \bm{\kappa}_{\mathrm{c}}(\bm{r})^{\ast} \bm{\kappa}_{\mathrm{c}}(\bm{r})^{\mathsf{T}} \mathrm{d}\bm{r} \bm{P}
\label{eq:wpm_W}
\end{align}
with 
\begin{align}
 \bm{P} := \left( \bm{K}_{\mathrm{c}} + \xi \bm{I} \right)^{-1}.
\label{eq:wpm_W_P}
\end{align}
The resulting cost function can be regarded as the weighted mean square error between the synthesized and desired pressures at the control points. Note that the weighting matrix $\bm{W}_{\mathrm{PM}}$ can be computed only with the positions of the control points and the target region $\Omega$. 

The optimization problem of the weighted pressure matching is formulated using the approximated cost function \eqref{eq:cost_wpm} as
\begin{align}
 \minimize_{\bm{d}\in\mathbb{C}^L} \left(\bm{Gd} - \bm{u}^{\mathrm{des}} \right)^{\mathsf{H}} \bm{W}_{\mathrm{PM}} \left(\bm{Gd} - \bm{u}^{\mathrm{des}} \right) + \lambda \|\bm{d}\|^2,
\label{eq:wpm_prob}
\end{align}
where $\lambda$ is the regularization parameter. This weighted least squares problem also has the closed-form solution as
\begin{align}
 \bm{d}_{\mathrm{WPM}} = \left( \bm{G}^{\mathsf{H}}\bm{W}_{\mathrm{PM}}\bm{G} + \lambda \bm{I} \right)^{-1} \bm{G}^{\mathsf{H}} \bm{W}_{\mathrm{PM}} \bm{u}_{\mathrm{des}}.
\label{eq:wpm_sol}
\end{align}
The weighted pressure matching enables the enhancement of the reproduction accuracy of pressure matching only by introducing the weighting matrix $\bm{W}_{\mathrm{PM}}$. This idea has already been applied in the context of the spatial active noise control~\cite{Ito:ICASSP2019,Koyama:IEEE_ACM_J_ASLP2021}. This interpolation-based sound field reproduction method is particularly effective when the region that the control points can be placed is limited.

\subsection{Weighted mode matching}
\label{sec:wmm}

Weighted mode matching is a method of solving the minimization problem of Eq.~\eqref{eq:cost} on the basis of the spherical wavefunction expansion of the sound field. The desired sound field $u_{\mathrm{des}}(\bm{r})$ and transfer function of the $l$th secondary source $g_l(\bm{r})$ are expanded around the expansion center $\bm{r}_{\mathrm{o}}$ as 
\begin{align}
 u_{\mathrm{des}}(\bm{r}) &= \sum_{\nu=0}^{\infty} \sum_{\mu=-\nu}^{\nu} \mathring{u}_{\mathrm{des},\nu,\mu}(\bm{r}_{\mathrm{o}}) \varphi_{\nu,\mu}(\bm{r}-\bm{r}_{\mathrm{o}}) \\
 g_l(\bm{r}) &= \sum_{\nu=0}^{\infty} \sum_{\mu=-\nu}^{\nu} \mathring{g}_{l.\nu,\mu} (\bm{r}_{\mathrm{o}}) \varphi_{\nu,\mu}(\bm{r}-\bm{r}_{\mathrm{o}}).
\label{eq:sphexp_ug}
\end{align}
By truncating the maximum order of the expansion in Eq.~\eqref{eq:sphexp_ug} up to $N_{\mathrm{tr}}$, we can approximate $u_{\mathrm{des}}$ and $\bm{g}(\bm{r})^{\mathsf{T}}$ as
\begin{align}
 u_{\mathrm{des}}(\bm{r}) &\approx \bar{\bm{\varphi}}(\bm{r})^{\mathsf{T}} \mathring{\bm{u}}^{\mathrm{des}} \\
 \bm{g}(\bm{r})^{\mathsf{T}} &\approx \bar{\bm{\varphi}}(\bm{r})^{\mathsf{T}} \mathring{\bm{G}}, 
\end{align}
where $\bar{\bm{\varphi}}(\bm{r})\in\mathbb{C}^{(N_{\mathrm{tr}}+1)^2}$, $\mathring{\bm{u}}^{\mathrm{des}}\in\mathbb{C}^{(N_{\mathrm{tr}}+1)^2}$, and $\mathring{\bm{G}}\in\mathbb{C}^{(N_{\mathrm{tr}}+1)^2 \times L}$ are the vectors and matrix consisting of $\varphi_{\nu,\mu}(\bm{r}-\bm{r}_{\mathrm{o}})$, $\mathring{u}_{\mathrm{des},\nu,\mu}(\bm{r}_{\mathrm{o}})$, and $\mathring{g}_{l,\nu,\mu}(\bm{r}_{\mathrm{o}})$, respectively. Thus, the cost function $J$ is approximated as
\begin{align}
 J &\approx \int_{\Omega} \left| \bar{\bm{\varphi}}(\bm{r})^{\mathsf{T}} \left( \mathring{\bm{G}}\bm{d} - \mathring{\bm{u}}^{\mathrm{des}} \right) \right|^2 \mathrm{d}\bm{r} \notag\\
 &= \left( \mathring{\bm{G}}\bm{d} - \mathring{\bm{u}}^{\mathrm{des}} \right)^{\mathsf{H}} \bm{W}_{\mathrm{MM}}  \left( \mathring{\bm{G}}\bm{d} - \mathring{\bm{u}}^{\mathrm{des}} \right),
\label{eq:cost_wmm}
\end{align}
where $\bm{W}_{\mathrm{MM}}\in\mathbb{C}^{(N_{\mathrm{tr}}+1)^2 \times (N_{\mathrm{tr}}+1)^2}$ is defined as
\begin{align}
 \bm{W}_{\mathrm{MM}} := \int_{\Omega} \bar{\bm{\varphi}}(\bm{r})^{\ast} \bar{\bm{\varphi}}(\bm{r})^{\mathsf{T}} \mathrm{d}\bm{r}.
\label{eq:weight_wmm}
\end{align}
As in the weighted pressure matching, the resulting cost function can be regarded as the weighted mean square error between synthesized and desired expansion coefficients around $\bm{r}_{\mathrm{o}}$. The weighting matrix $\bm{W}_{\mathrm{MM}}$ can be computed only by using the spherical wavefunctions and target region $\Omega$. In a 2D sound field, the spherical wavefunctions in the integrand are replaced with the cylindrical wavefunctions~\cite{Ueno:IEEE_SPL2018}. When $\mathring{\bm{u}}^{\mathrm{des}}$ and $\mathring{\bm{G}}$ are obtained from measurements, for example, to reproduce a captured sound field and/or to compensate for reverberation in the transfer functions of secondary sources, sound field capturing methods such as the infinite-dimensional harmonic analysis introduced in Section~\ref{sec:sf_est} can be applied. 

The optimization problem of the weighted mode matching is formulated using the approximated cost function $J$ in Eq.~\eqref{eq:cost_wmm} as
\begin{align}
 \minimize_{\bm{d}\in\mathbb{C}^L} \left( \mathring{\bm{G}}\bm{d} - \mathring{\bm{u}}^{\mathrm{des}} \right)^{\mathsf{H}} \bm{W}_{\mathrm{MM}}  \left( \mathring{\bm{G}}\bm{d} - \mathring{\bm{u}}^{\mathrm{des}} \right) + \gamma \|\bm{d}\|^2,
\label{eq:wmm_prob}
\end{align}
where $\gamma$ is the regularization parameter. Again, this weighted least squares problem can be solved as
\begin{align}
 \bm{d}_{\mathrm{WMM}} = \left( \mathring{\bm{G}}^{\mathsf{H}}\bm{W}_{\mathrm{MM}}\mathring{\bm{G}} + \gamma \bm{I} \right)^{-1} \mathring{\bm{G}}^{\mathsf{H}} \bm{W}_{\mathrm{MM}} \mathring{\bm{u}}^{\mathrm{des}}.
\label{eq:wmm_sol}
\end{align}
The weights for each expansion coefficient are determined by the weighting matrix $\bm{W}_{\mathrm{MM}}$. When $\bm{W}_{\mathrm{MM}}$ is the identity matrix, Eq.~\eqref{eq:wmm_sol} corresponds to the driving signal of standard mode matching.
\begin{align}
 \bm{d}_{\mathrm{MM}} = \left( \mathring{\bm{G}}^{\mathsf{H}}\mathring{\bm{G}} + \gamma \bm{I} \right)^{-1} \mathring{\bm{G}}^{\mathsf{H}} \mathring{\bm{u}}^{\mathrm{des}}
\label{eq:mm_sol}
\end{align}
In the mode matching, the appropriate setting of the truncation order $N_{\mathrm{tr}}$ for the spherical wavefunction expansion is necessary. When the target region $\Omega$ is a spherical region of radius $R$, $N_{\mathrm{tr}}=\lceil kR \rceil$ is empirically known to be a proper truncation criterion; however, when $\Omega$ is not spherical, the appropriate setting of $N_{\mathrm{tr}}$ is not simple. In particular, the target region of the sound field reproduction is sometimes set to be around a horizontal plane because listeners can be considered not to move largely in the vertical directions.

\subsection{Relationship between weighted pressure and mode matching}
\label{sec:rel}

As discussed in Sections~\ref{sec:wpm} and \ref{sec:wmm}, the weighted pressure and mode matching can be regarded as a generalization of pressure and mode matching. Furthermore, the weighted pressure matching can be regarded as a special case of the weighted mode matching. Suppose that the expansion coefficients $\mathring{\bm{u}}^{\mathrm{des}}$ and $\mathring{\bm{G}}$ are estimated from the pressure observations at the control points $\left\{ \bm{r}_{\mathrm{c},n} \right\}_{n=1}^N$. On the basis of infinite-dimensional harmonic analysis in Section~\ref{sec:sf_est}, $\mathring{\bm{u}}^{\mathrm{des}}$ and $\mathring{\bm{G}}$ are estimated as
\begin{align}
 \hat{\mathring{\bm{u}}}^{\mathrm{des}}  &= \bm{\Xi}_{\mathrm{c}}(\bm{r}_{\mathrm{o}}) (\bm{\Psi}_{\mathrm{c}} + \xi \bm{I})^{-1} \bm{u}^{\mathrm{des}} \notag\\
 \hat{\mathring{\bm{G}}} &= \bm{\Xi}_{\mathrm{c}}(\bm{r}_{\mathrm{o}}) (\bm{\Psi}_{\mathrm{c}} + \xi \bm{I})^{-1} \bm{G},
\end{align} 
where $\bm{\Xi}_{\mathrm{c}}$ and $\bm{\Psi}_{\mathrm{c}}$ are the matrices defined in Eqs.~\eqref{eq:est_xi} and \eqref{eq:est_psi} with the control positions $\{\bm{r}_{\mathrm{c},n}\}_{n=1}^N$, respectively. Therefore, the cost function $J$ of the weighted mode matching becomes
\begin{align}
 J &\approx \left( \hat{\mathring{\bm{G}}}\bm{d} - \hat{\mathring{\bm{u}}}^{\mathrm{des}} \right)^{\mathsf{H}} \bm{W}_{\mathrm{MM}}  \left( \hat{\mathring{\bm{G}}}\bm{d} - \hat{\mathring{\bm{u}}}^{\mathrm{des}} \right) \notag\\
 &= \left(\bm{Gd} - \bm{u}^{\mathrm{des}}\right)^{\mathsf{H}} \bm{Q}^{\mathsf{H}} \bm{\Xi}_{\mathrm{c}}(\bm{r}_{\mathrm{o}})^{\mathsf{H}} \bm{W}_{\mathrm{MM}} \bm{\Xi}_{\mathrm{c}}(\bm{r}_{\mathrm{o}}) \bm{Q} \left(\bm{Gd} - \bm{u}^{\mathrm{des}}\right),
\end{align}
where 
\begin{align}
 \bm{Q} := (\bm{\Psi}_{\mathrm{c}} + \xi \bm{I})^{-1}.
\end{align}
Since the observations at the control points are assumed to be pressure, i.e., ominidirectional microphone measurements, $\bm{\Psi}_{\mathrm{c}}$ is equivalent to $\bm{K}_{\mathrm{c}}$, thus $\bm{Q}=\bm{P}$. Moreover, $\bm{\Xi}_{\mathrm{c}}(\bm{r}_{\mathrm{o}})^{\mathsf{H}}\bm{W}_{\mathrm{MM}}\bm{\Xi}_{\mathrm{c}}(\bm{r}_{\mathrm{o}})$ is calculated as
\begin{align}
& \bm{\Xi}_{\mathrm{c}}(\bm{r}_{\mathrm{o}})^{\mathsf{H}}\bm{W}_{\mathrm{MM}}\bm{\Xi}_{\mathrm{c}}(\bm{r}_{\mathrm{o}}) \notag\\
&= \quad \int_{\Omega} \left( \bm{\varphi}(\bm{r}-\bm{r}_{\mathrm{o}})^{\mathsf{T}}\bm{\Xi}_{\mathrm{c}}(\bm{r}_{\mathrm{o}}) \right)^{\mathsf{H}} \left( \bm{\varphi}(\bm{r}-\bm{r}_{\mathrm{o}})^{\mathsf{T}}\bm{\Xi}_{\mathrm{c}}(\bm{r}_{\mathrm{o}}) \right) \mathrm{d} \bm{r} \notag\\
&= \quad \int_{\Omega} \bm{\kappa}_{\mathrm{c}}(\bm{r})^{\ast} \bm{\kappa}_{\mathrm{c}}(\bm{r})^{\mathsf{T}} \mathrm{d}\bm{r},
\end{align}
because 
\begin{align}
 \bm{\varphi}(\bm{r}-\bm{r}_{\mathrm{o}})^{\mathsf{T}}\bm{\Xi}_{\mathrm{c}}(\bm{r}_{\mathrm{o}}) &= 
\begin{bmatrix}
 \bm{\varphi}(\bm{r}-\bm{r}_{\mathrm{1}})^{\mathsf{T}}\bm{c}_1, & \ldots, & \bm{\varphi}(\bm{r}-\bm{r}_{\mathrm{N}})^{\mathsf{T}}\bm{c}_N
\end{bmatrix} \notag\\
&= 
\begin{bmatrix}
 j_0(k\|\bm{r}-\bm{r}_1\|), & \ldots, & j_0(k\|\bm{r}-\bm{r}_N\|)
\end{bmatrix}.
\end{align}
Here, property \eqref{eq:trans_prop3} is used. Note that $\{\bm{c}_n\}_{n=1}^{N}$ is obtained as Eq.~\eqref{eq:c_omni}. In summary, when the expansion coefficients $\mathring{\bm{u}}^{\mathrm{des}}$ and $\mathring{\bm{G}}$ in the weighted mode matching are obtained by infinite-dimensional harmonic analysis from the pressure observations at the control points $\bm{u}^{\mathrm{des}}$ and $\bm{G}$, the weighted mode matching corresponds to the weighted pressure matching. 

\section{EXPERIMENTS}
\label{sec:exp}

We conducted experiments to compare pressure matching, weighted pressure matching, mode matching, and weighted mode matching, which are hereafter denoted as PM, WPM, MM, and WMM, respectively. First, we show numerical simulation results. Then, experimental results obtained using real data are demonstrated. 

\subsection{Numerical simulation}
\label{sec:exp_sim}

\begin{figure}[t]
\centering
\includegraphics[width=65mm]{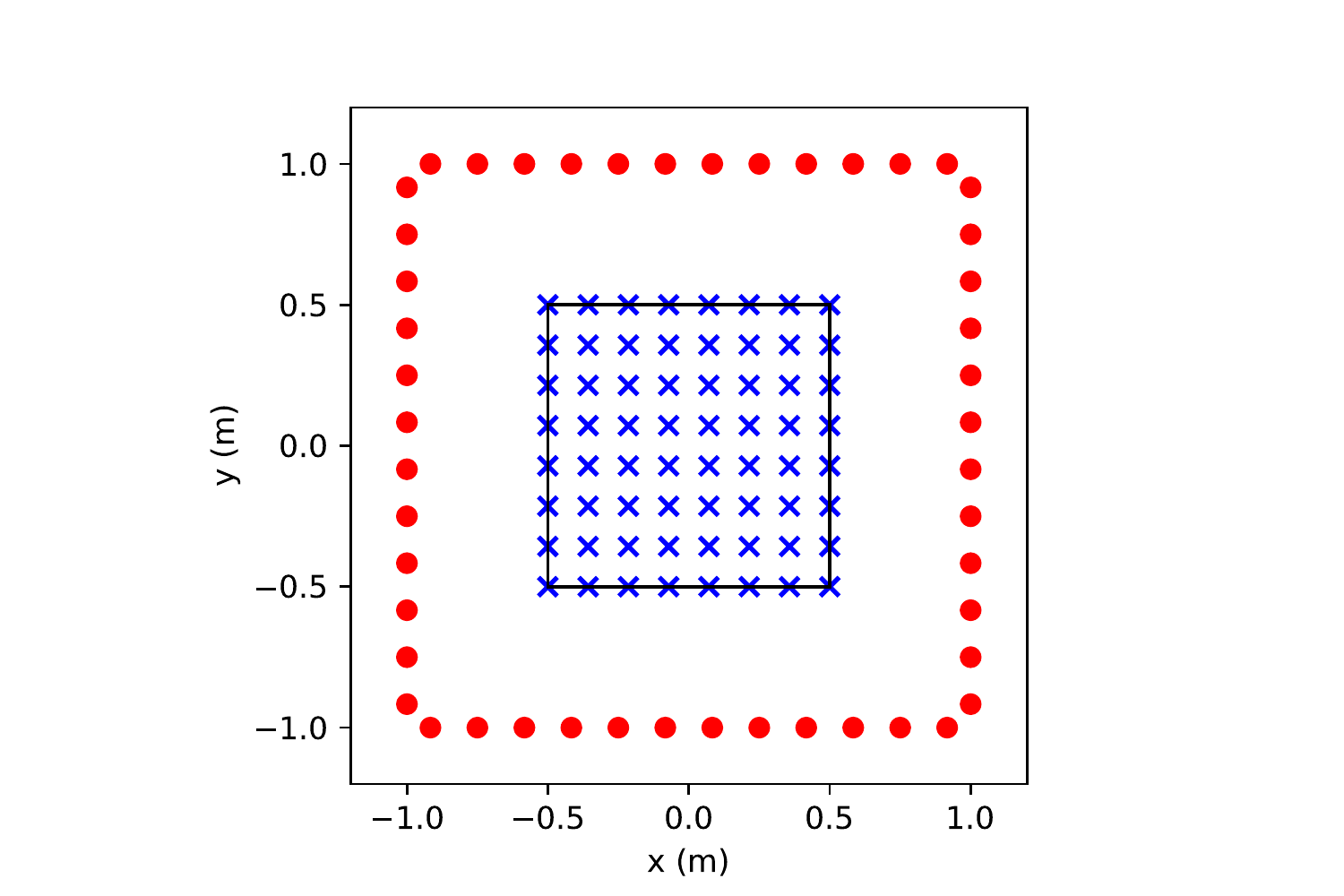}
\caption{Experimental setup for numerical simulation. The target region was set as a 2D square region. Red dots and blue crosses indicate loudspeakers and control points, respectively.}
\label{fig:exp_cond_sim}
\end{figure}

The reproduction performances of the four methods are evaluated by numerical simulation in a 3D free field. Figure~\ref{fig:exp_cond_sim} shows the experimental setup. A target region, loudspeakers, and control points were set on the $x$-$y$-plane at $z=0$. Forty-eight loudspeakers were regularly placed along the border of a square with dimensions $2.0~\mathrm{m} \times 2.0~\mathrm{m}$. The target region $\Omega$ was set as a 2D square region of $1.0~\mathrm{m} \times 1.0~\mathrm{m}$ at $z=0$. The centers of these squares were at the origin. Thirty-six control points were regularly placed over the target region. In Fig.~\ref{fig:exp_cond_sim}, the loudspeakers and control points are indicated by red dots and blue crosses, respectively. Each loudspeaker was assumed to be a point source. The desired sound field was a single plane wave, whose propagation direction was $(\theta,\phi)=(\pi/2, \pi/4)~\mathrm{rad}$. 

In PM and WPM, $\bm{u}^{\mathrm{des}}$ and $\bm{G}$ in Eqs.~\eqref{eq:pm_sol} and \eqref{eq:wpm_sol} were given as pressure values at the control points. The expansion coefficients $\mathring{\bm{G}}$ in Eqs.~\eqref{eq:mm_sol} and \eqref{eq:wmm_sol} were estimated up to the maximum order $N_{\mathrm{tr}}$ from $\bm{G}$ by infinite-dimensional harmonic analysis \eqref{eq:est} in the mode and weighted mode matching. The desired expansion coefficients $\mathring{\bm{u}}^{\mathrm{des}}$ were analytically given up to $N_{\mathrm{tr}}$. In MM, the truncation order was determined as $N_{\mathrm{tr}}=\lceil k R \rceil$, where $R$ was set to $0.5\sqrt{2}~\mathrm{m}$ to cover the target region. Furthermore, to enhance the reproduction accuracy on the $x$-$y$-plane at $z=0$, the coefficients of $\nu=|\mu|$ were only used~\cite{Travis:AmbiSymp2009}. The truncation order $N_{\mathrm{tr}}$ for WMM was set to $30$, which is sufficiently larger than the maximum required order of MM. The regularization parameters in Eqs.~\eqref{eq:pm_sol}, \eqref{eq:wpm_sol}, \eqref{eq:mm_sol}, and \eqref{eq:wmm_sol} were determined at each frequency as $\sigma_{\max}^2(\bm{A}) \times 10^{-3}$, where $\sigma_{\max}^2(\bm{A})$ is the maximum eigenvalue of the matrix to be inverted $\bm{A}$. Therefore, $\bm{A}$ is $\bm{G}^{\mathsf{H}}\bm{G}$, $\bm{G}^{\mathsf{H}}\bm{W}_{\mathrm{PM}}\bm{G}$, $\mathring{\bm{G}}^{\mathsf{H}}\mathring{\bm{G}}$, and $\mathring{\bm{G}}^{\mathsf{H}}\bm{W}_{\mathrm{MM}}\mathring{\bm{G}}$ in PM, WPM, MM, and WMM, respectively. The parameter $\xi$ in Eqs.~\eqref{eq:wpm_W_P} and \eqref{eq:est} was set as $\sigma_{\max}(\bm{K})\times 10^{-3}$ at each frequency.

\begin{figure}[t]
\centering
\includegraphics[width=85mm]{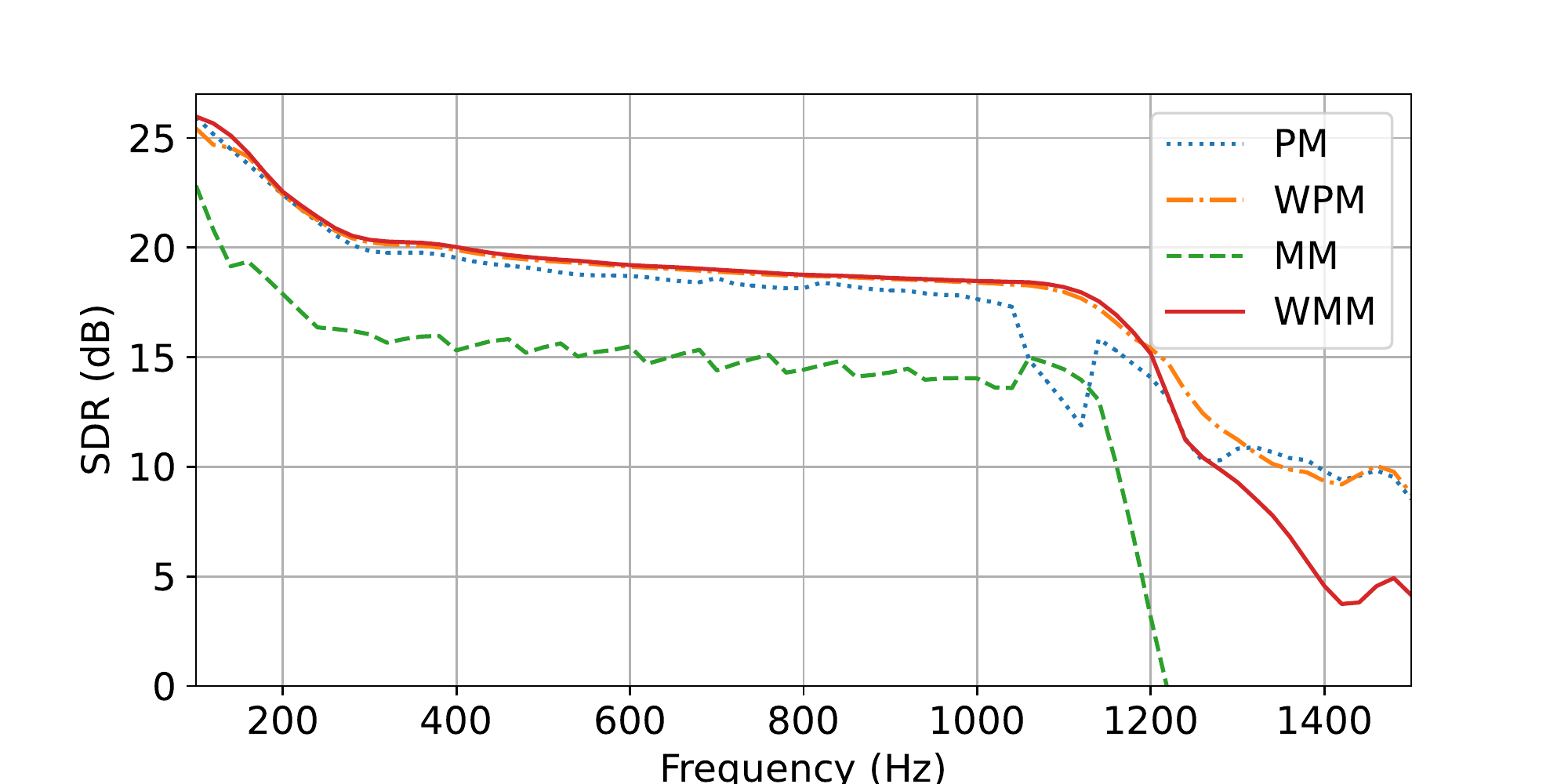}
\caption{$\mathrm{SDR}$ with respect to frequency.}
\label{fig:sdr_freq}
\end{figure}

\begin{figure}[t]
\centering
\subfloat[PM]{\includegraphics[height=30mm,clip]{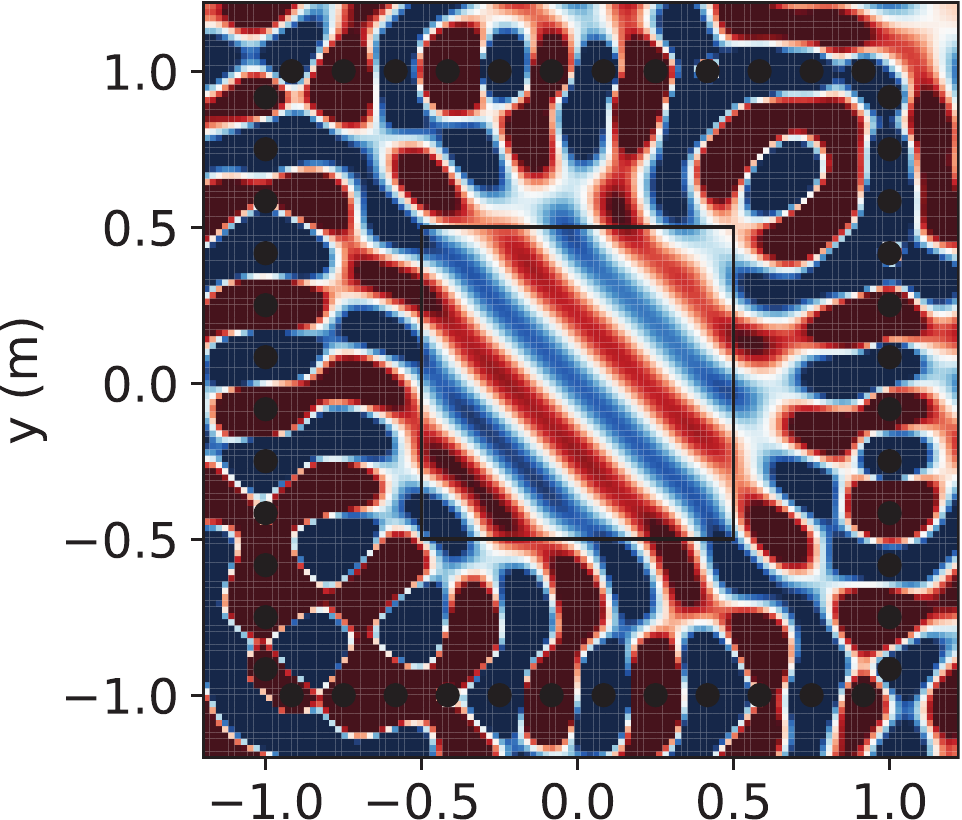} \hspace{10pt}}%
\subfloat[WPM]{\includegraphics[height=30mm,clip]{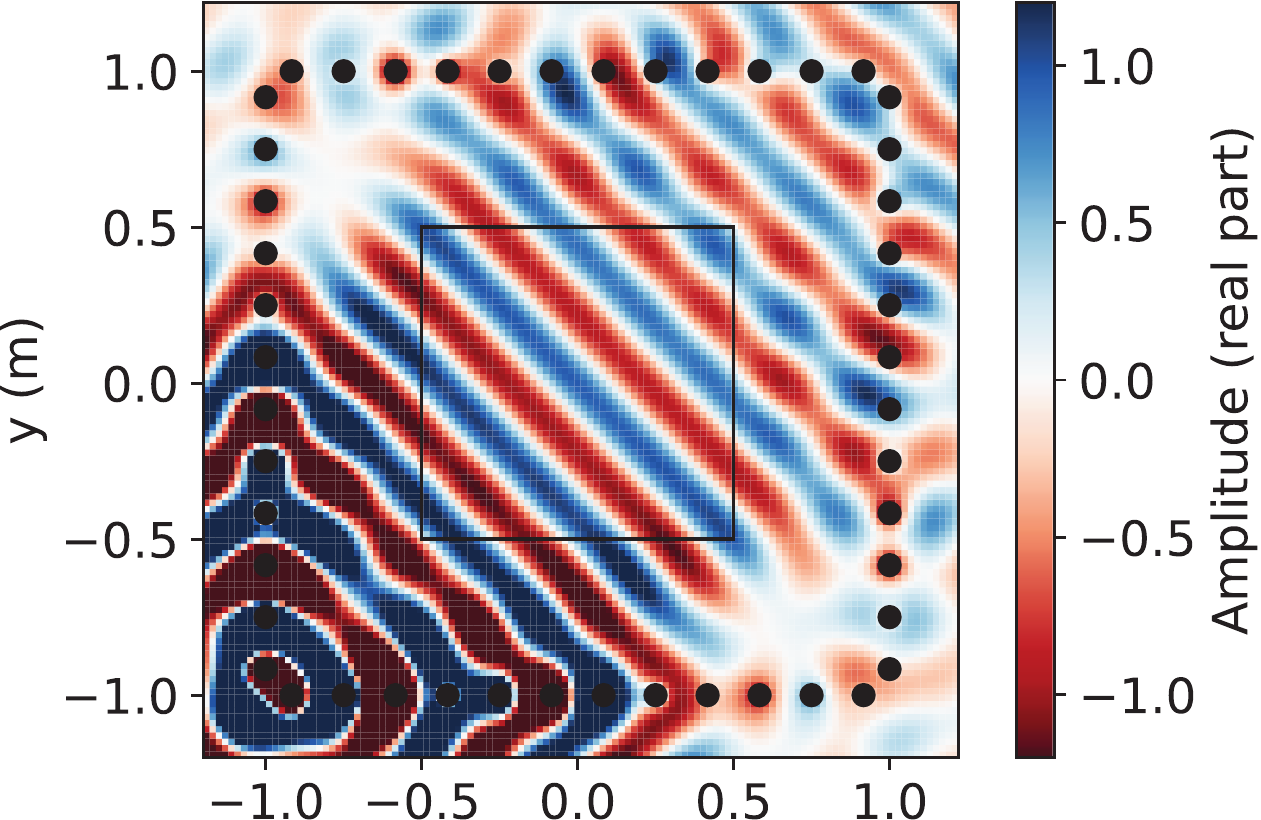}}\\
\subfloat[MM]{\includegraphics[height=30mm,clip]{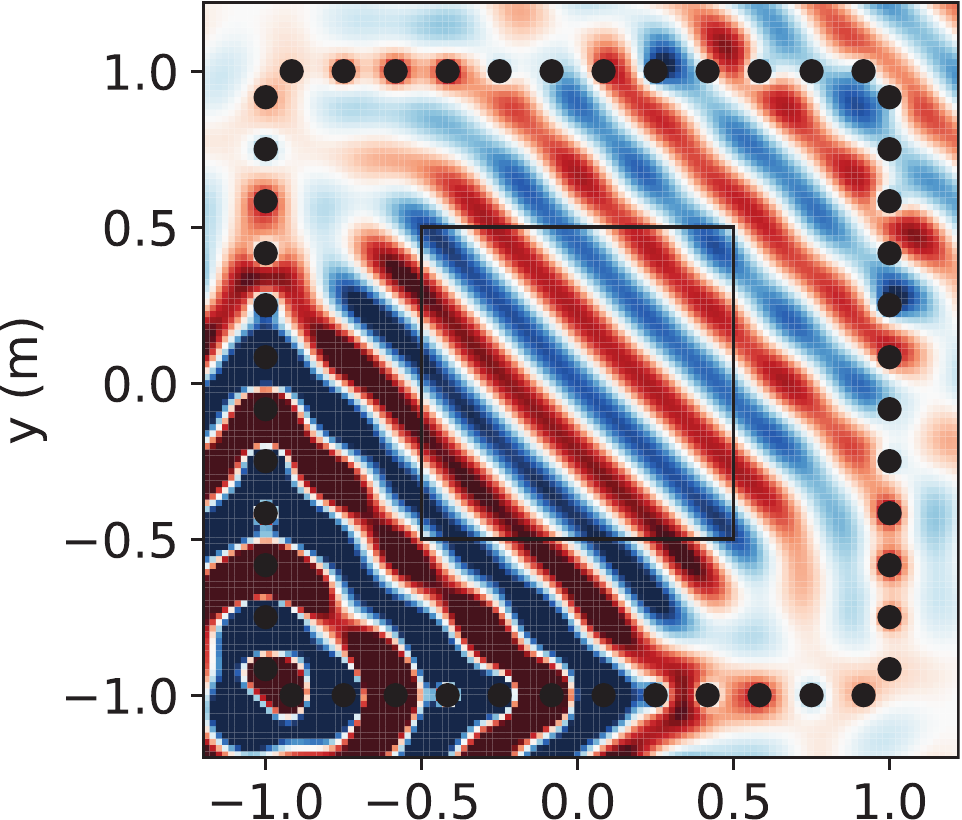} \hspace{10pt}}%
\subfloat[WMM]{\includegraphics[height=30mm,clip]{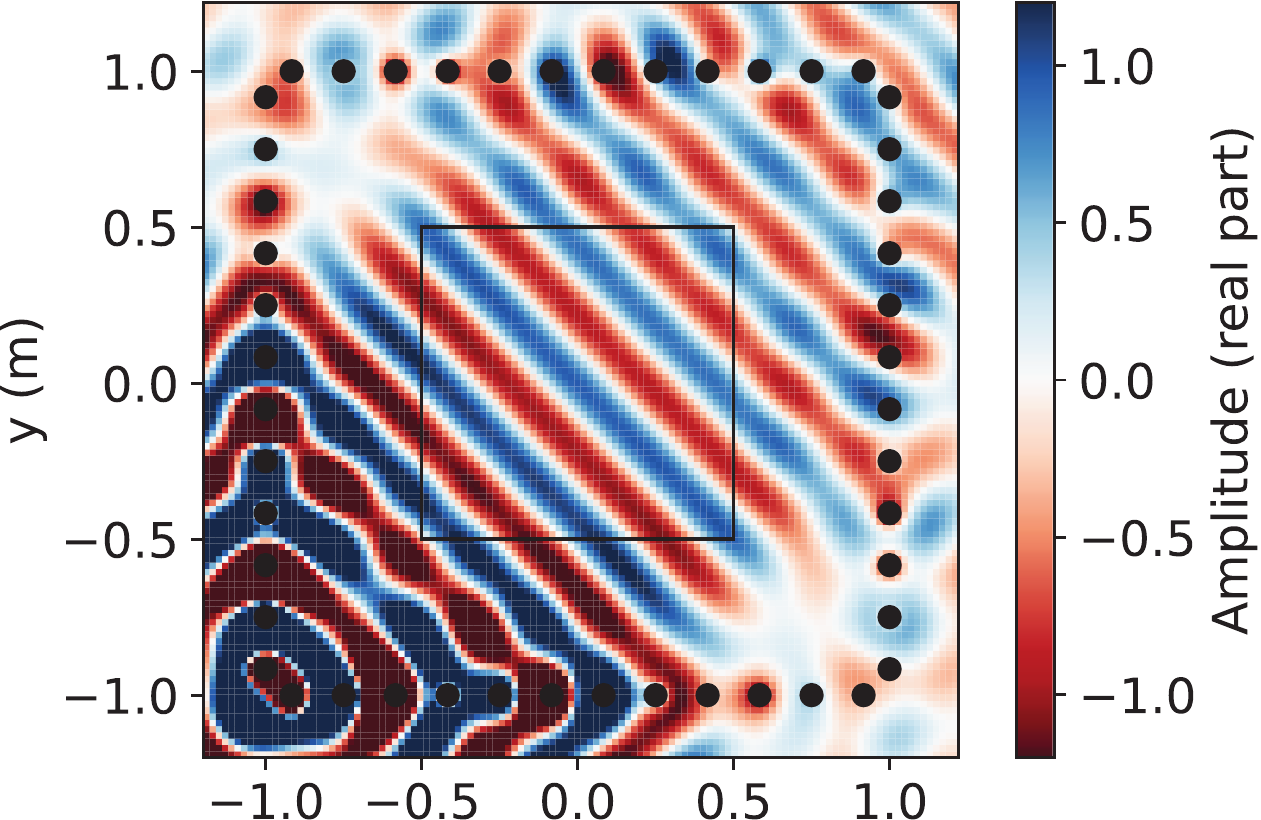}}
\caption{Reproduced pressure distribution at $1100~\mathrm{Hz}$. $\mathrm{SDR}$s of PM, WPM, MM, and WMM were $12.9$, $18.0$, $14.4$, and $18.2~\mathrm{dB}$, respectively.}
\label{fig:dist_freq}
\centering
\subfloat[PM]{\includegraphics[height=30mm]{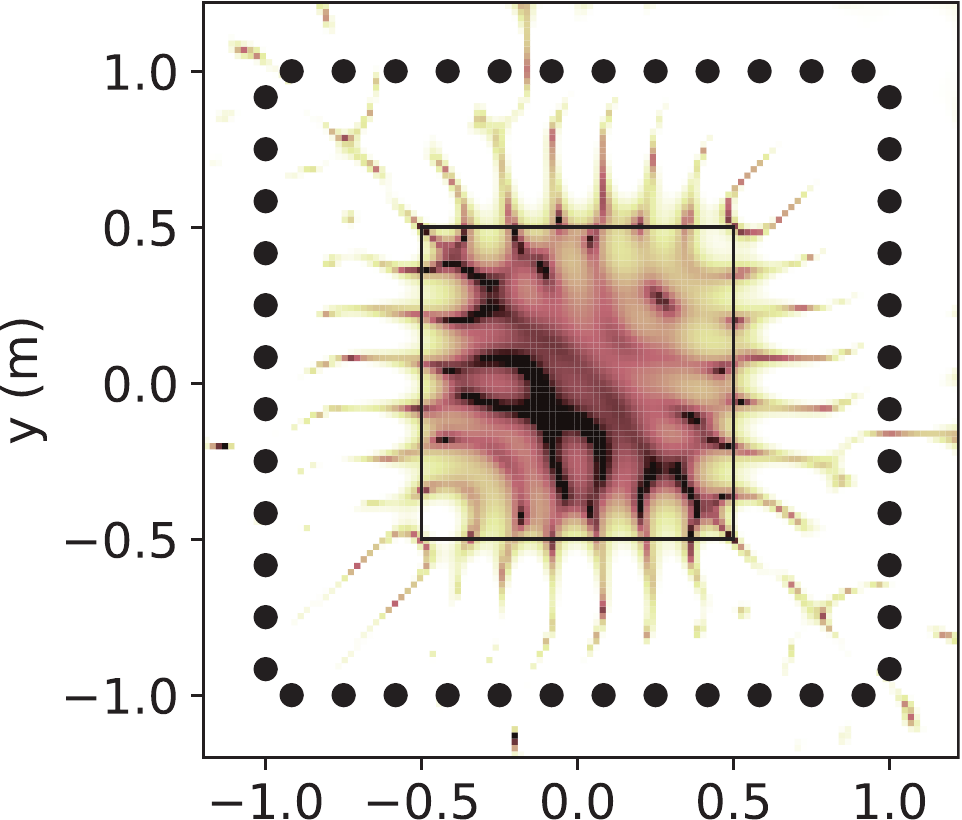} \hspace{10pt}}%
\subfloat[WPM]{\includegraphics[height=30mm]{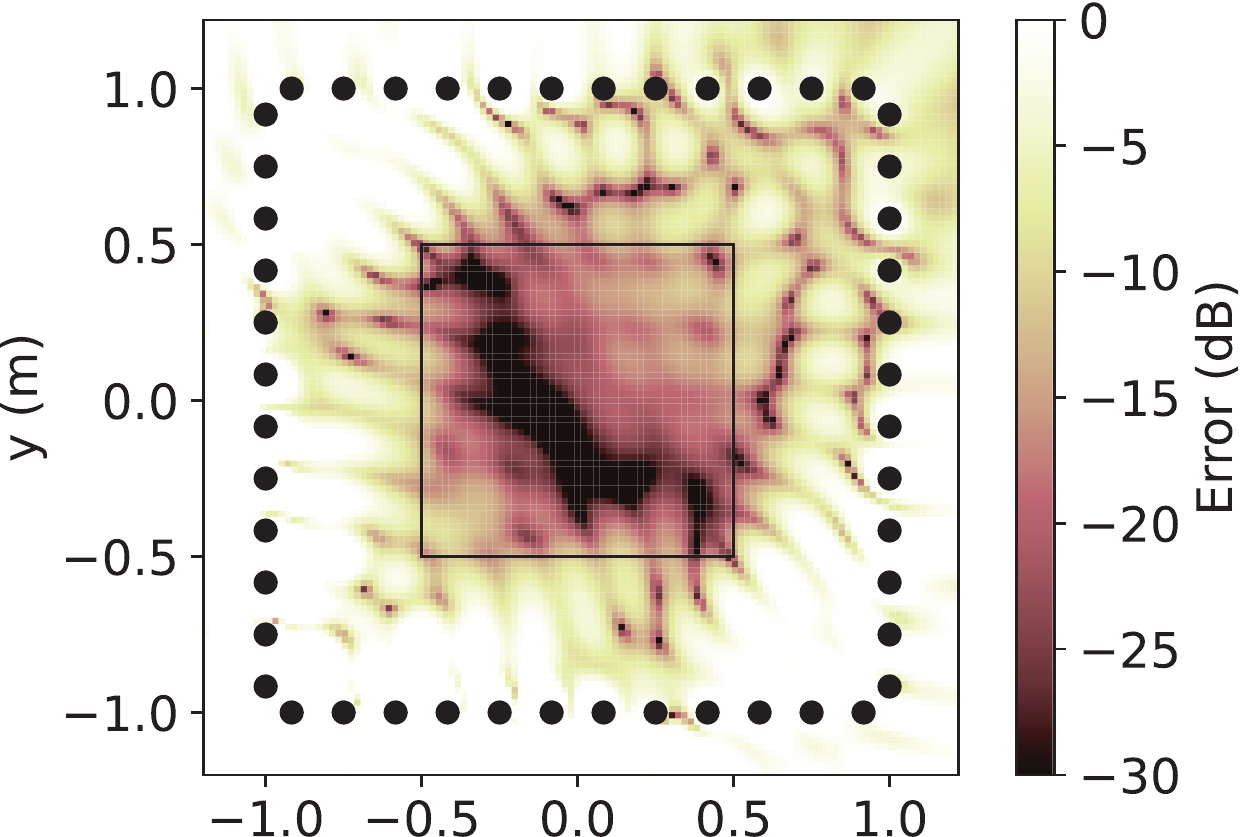}}\\
\subfloat[MM]{\includegraphics[height=30mm]{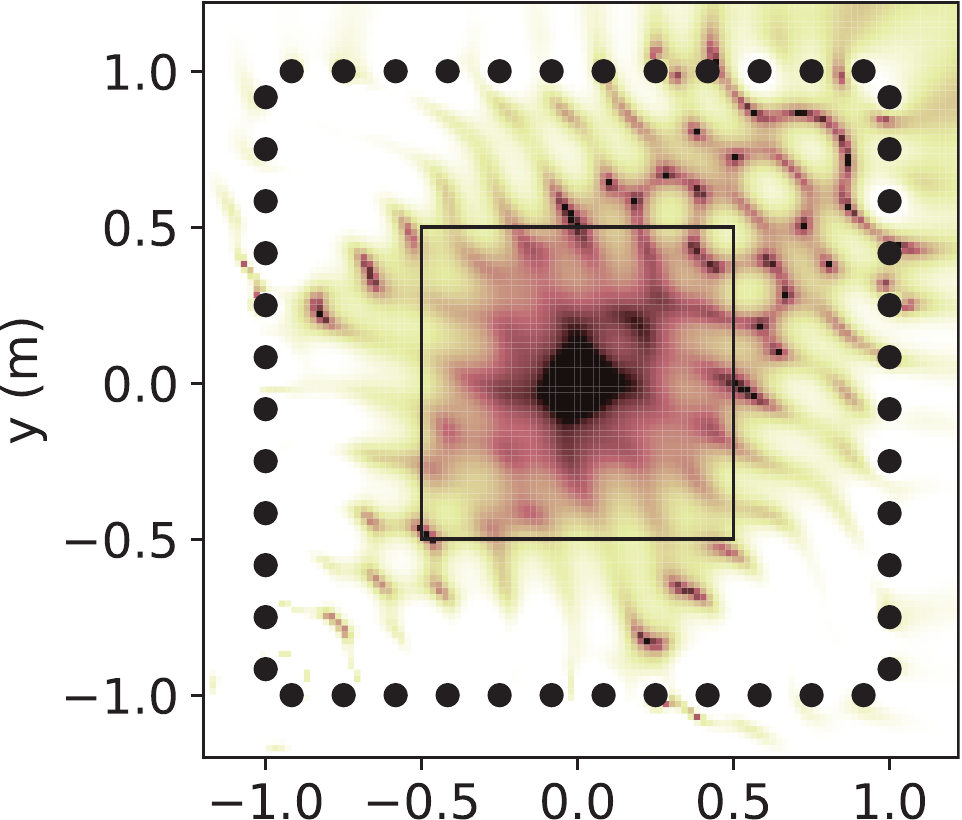} \hspace{10pt}}%
\subfloat[WMM]{\includegraphics[height=30mm]{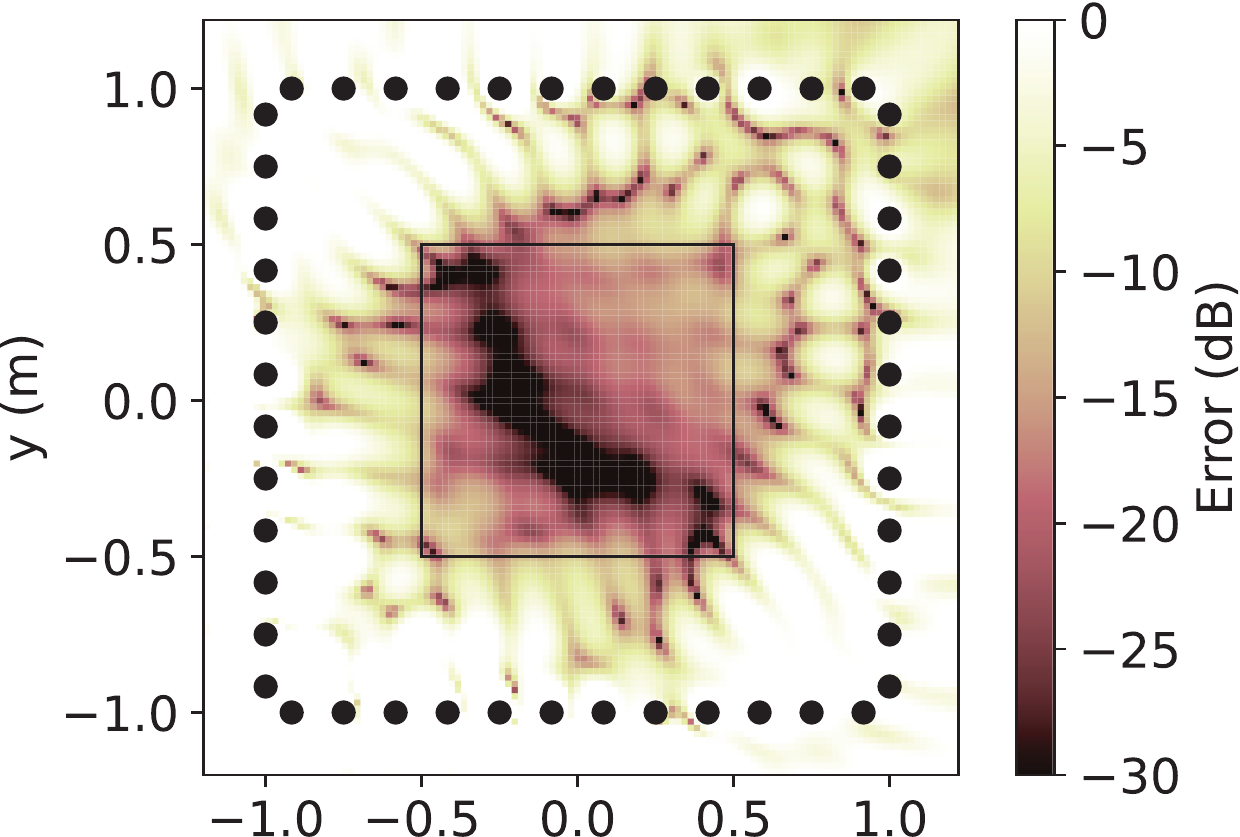}}
\caption{Square error distribution at $1100~\mathrm{Hz}$.}
\label{fig:err_freq}
\end{figure}

For evaluation measure in the frequency domain, we define the signal-to-distortion ratio ($\mathrm{SDR}$) as
\begin{align}
 \mathrm{SDR}(\omega) = \frac{\int_{\Omega} |u_{\mathrm{des}}(\bm{r},\omega)|^2 \mathrm{d}\bm{r}}{\int_{\Omega} |u_{\mathrm{syn}}(\bm{r},\omega) - u_{\mathrm{des}}(\bm{r},\omega)|^2 \mathrm{d}\bm{r}},
\end{align}
where the integration was computed at the evaluation points. The evaluation points were obtained by regularly discretizing the target region every $0.02~\mathrm{m}$.

The $\mathrm{SDR}$ with respect to the frequency is plotted from $100~\mathrm{Hz}$ to $1500~\mathrm{Hz}$ in Fig.~\ref{fig:sdr_freq}. The $\mathrm{SDR}$s of MM were smaller than those of the other three methods below $1000~\mathrm{Hz}$. This can be considered to be due to the empirical truncation and weighting for the expansion coefficients in MM. Note that the reproduction accuracy further deteriorated when all the expansion coefficients up to the truncation order were used without the extraction of $\nu=|\mu|$. The other three methods, PM, WPM, and WMM, achieved high reproduction accuracy. However, the $\mathrm{SDR}$s of PM sharply decreased above $1000~\mathrm{Hz}$. The $\mathrm{SDR}$s of WPM and WMM were slightly higher than those of PM below $1000~\mathrm{Hz}$, and they were maintained high up to $1100~\mathrm{Hz}$. Furthermore, the plots of WPM and WMM almost overlapped below $1200~\mathrm{Hz}$ because of the equivalence between the two methods except the setting of the desired sound field, i.e., the desired pressures at the control points or desired expansion coefficients.

As an example, the synthesized pressure distribution of each method at $1100~\mathrm{Hz}$ is shown in Fig.~\ref{fig:dist_freq}. Figure~\ref{fig:err_freq} is the square error distribution of each method at $1100~\mathrm{Hz}$. In WPM and WMM, the error was particularly small around a line in the target region. This is due to the 2D placement of the loudspeakers in 3D space. The amplitude of the synthesized sound field in PM was high outside the target region. In MM, the region of small reproduction error was limited around the center of the target region. The $\mathrm{SDR}$s at this frequency were $12.9$, $18.0$, $14.4$, and $18.2~\mathrm{dB}$ for PM, WPM, MM, and WMM, respectively.


\begin{figure}[t]
\centering
\includegraphics[width=85mm]{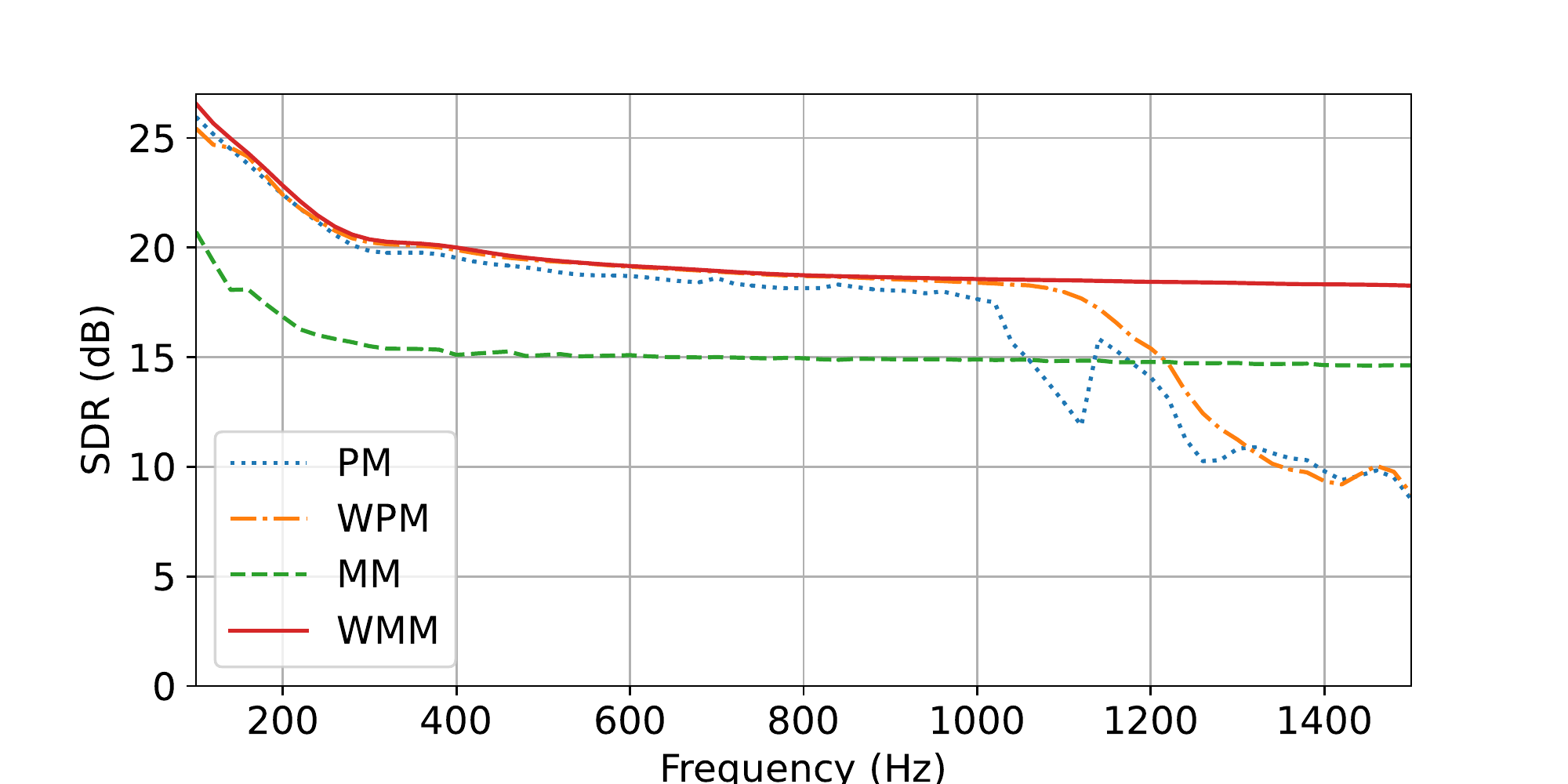}
\caption{$\mathrm{SDR}$ with respect to frequency when true expansion coefficients were used in MM and WMM.}
\label{fig:sdr_truecoef}
\end{figure}

\begin{figure}[t]
\centering
\includegraphics[width=85mm]{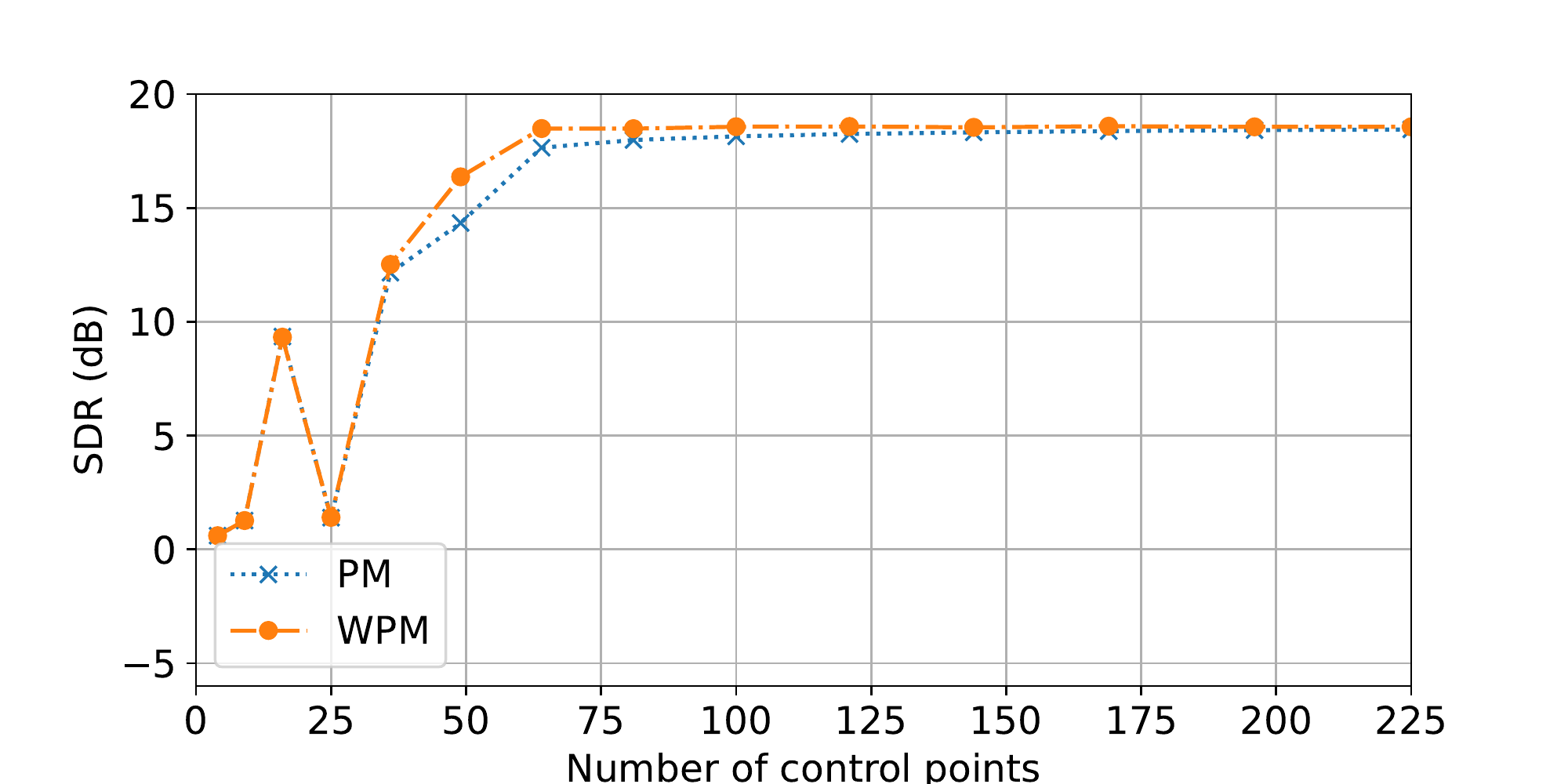}
\caption{$\mathrm{SDR}$ with respect to number of control points at $1000~\mathrm{Hz}$.}
\label{fig:sdr_numcp}
\end{figure}

\begin{figure}[t]
\centering
\includegraphics[width=80mm]{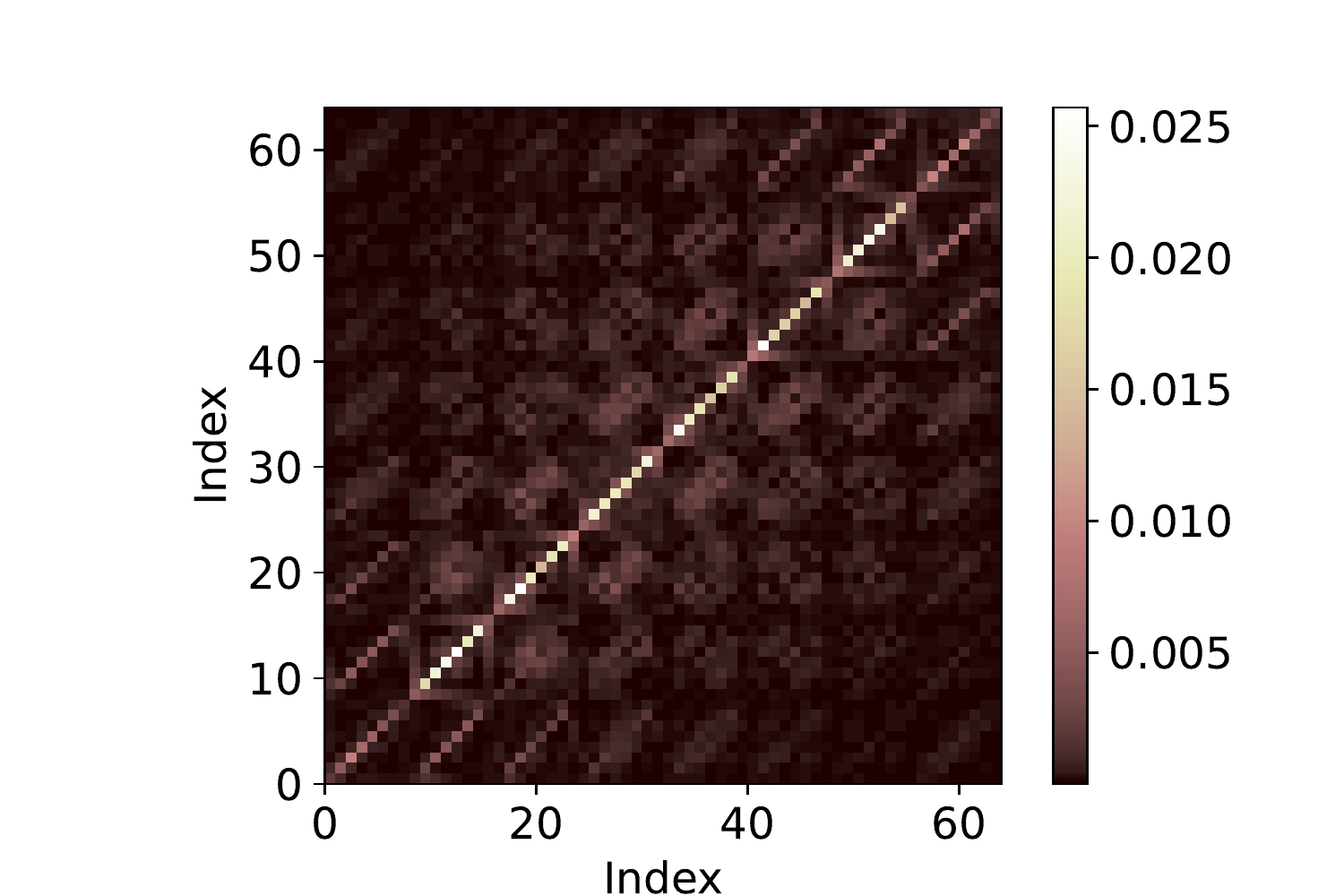}
\caption{Absolute value of weighting matrices of WPM $|\bm{W}_{\mathrm{PM}}|$ ($M=64$) at $1000~\mathrm{Hz}$.}
\label{fig:weight_wpm}
\end{figure}

\begin{figure}[t]
\centering
\includegraphics[width=85mm]{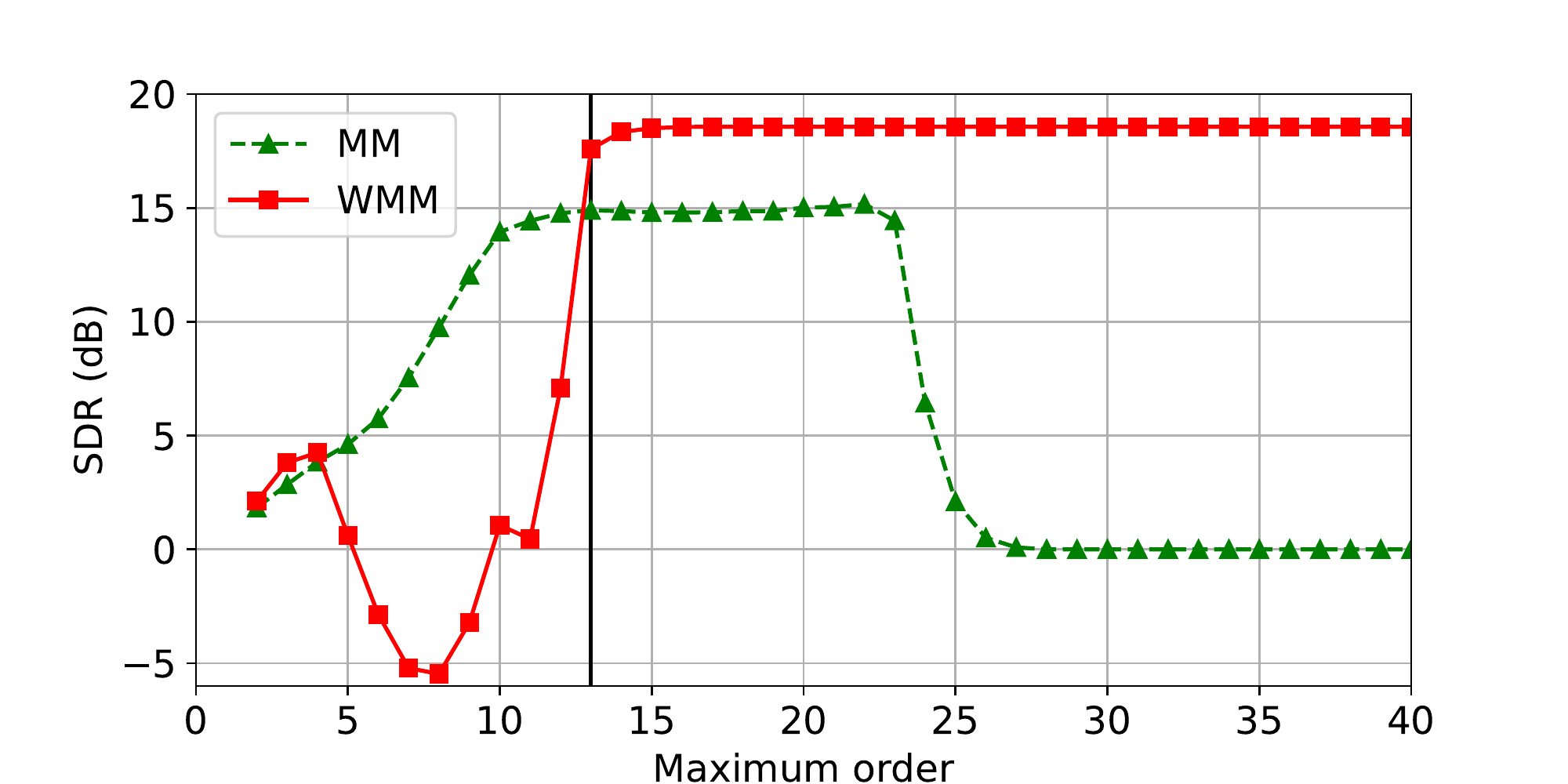}
\caption{$\mathrm{SDR}$ with respect to maximum order of spherical wavefunctions $N_{\mathrm{tr}}$ at $1000~\mathrm{Hz}$. Black line indicates the order of $\lceil kR \rceil$.}
\label{fig:sdr_order}
\end{figure}

\begin{figure}[t]
\centering
\subfloat[$|\bm{W}_{\mathrm{MM}}|$ ($N_{\mathrm{tr}}=7$)]{\includegraphics[width=80mm]{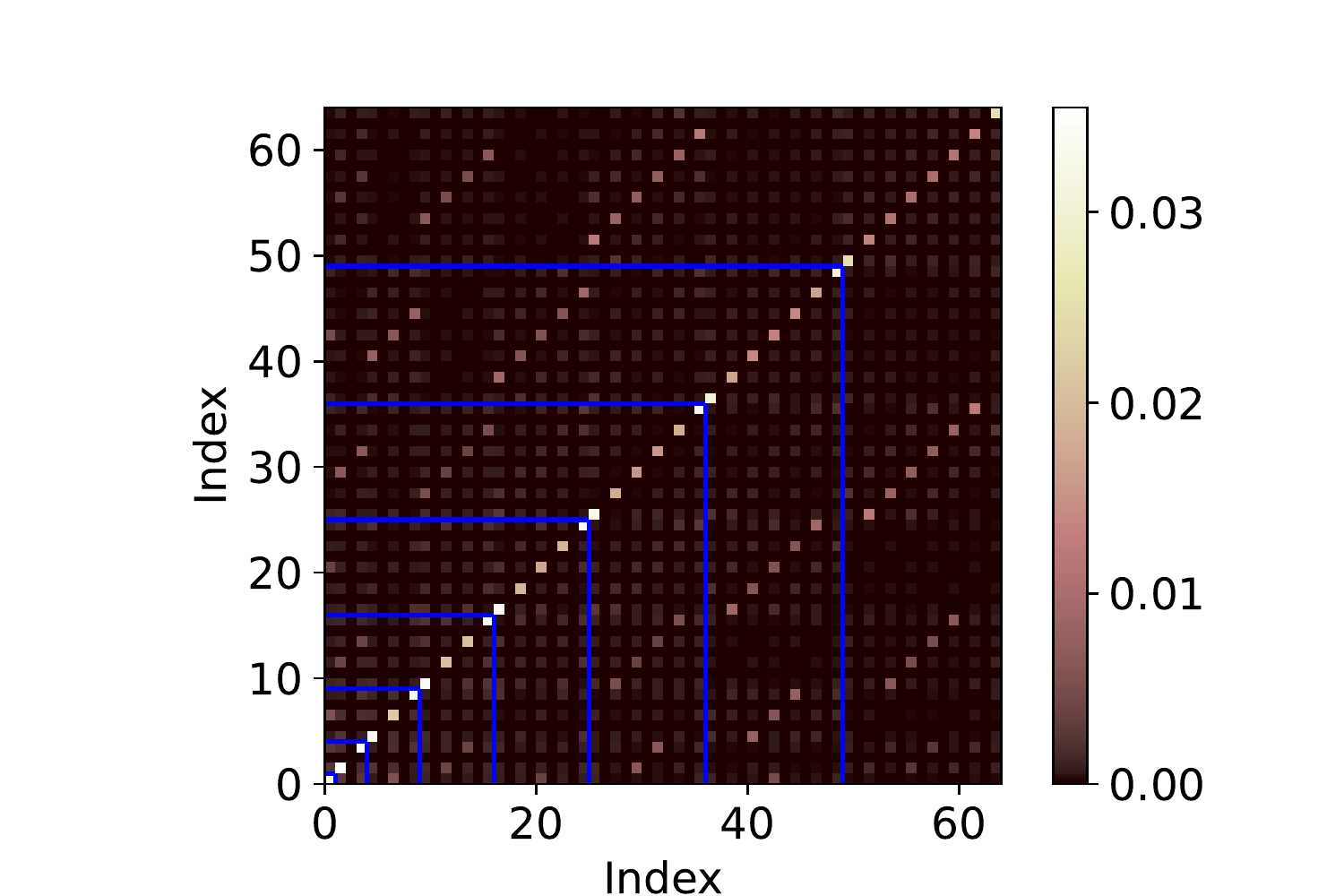}}\label{fig:weight_wmm_index}\\
\vspace{-20pt}
\subfloat[Diagonal elements of $|\bm{W}_{\mathrm{MM}}|$ ($N_{\mathrm{tr}}=18$)]{\includegraphics[width=90mm]{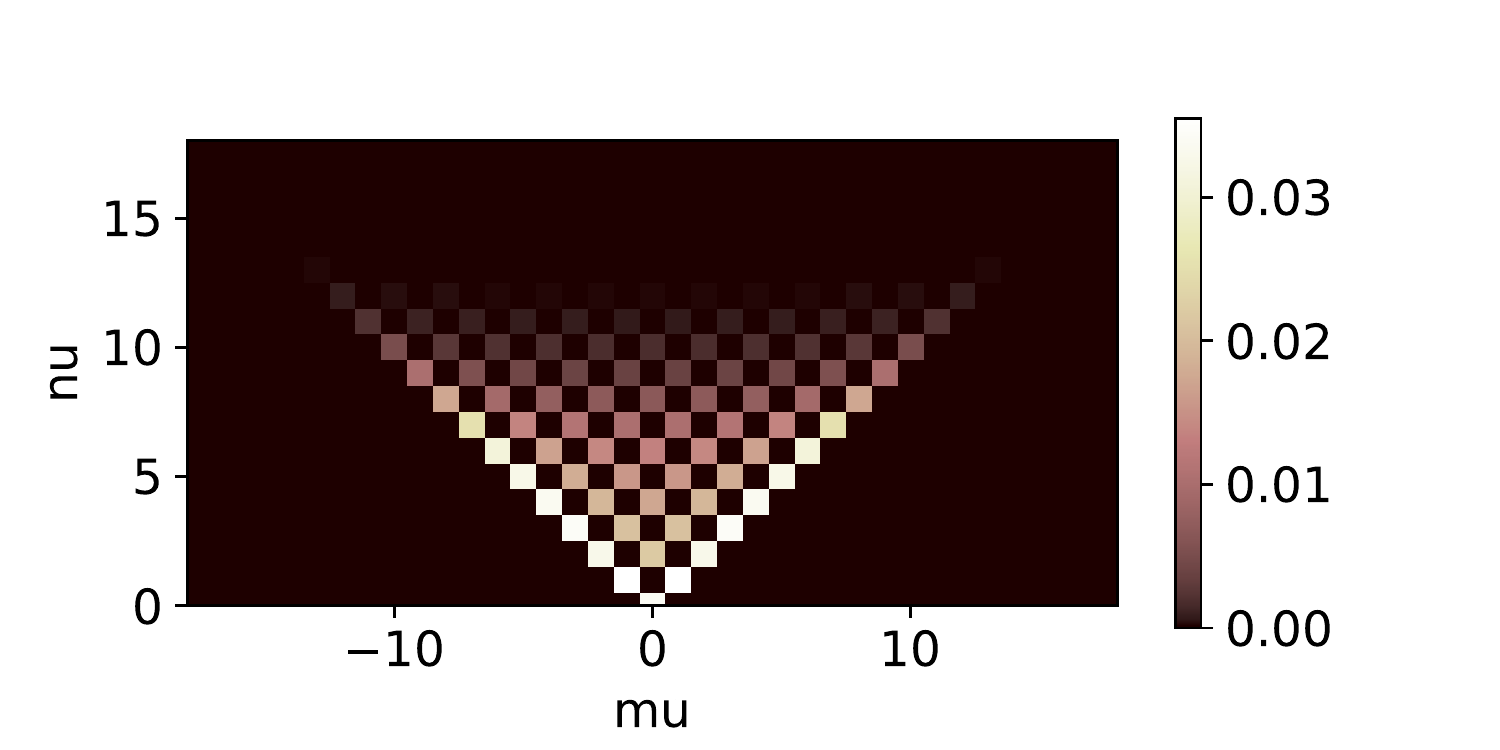}}\label{fig:weight_wmm_nm}
\caption{Absolute value of weighting matrices of WMM $|\bm{W}_{\mathrm{MM}}|$ and its diagonal elements sorted with respect to $\nu$ and $\mu$ at $1000~\mathrm{Hz}$. Blue lines in (a) indicate the range of the indexes of the same $\nu$.}
\label{fig:weight_wmm}
\end{figure}

Next, we consider the case that the expansion coefficients of the transfer functions $\mathring{\bm{G}}$ are also analytically given in MM and WMM to investigate the difference between WPM and WMM. The other settings were the same as the previous ones. Figure~\ref{fig:sdr_truecoef} shows the $\mathrm{SDR}$ with respect to the frequency. Note that the results of PM and WPM are the same as those in Fig.~\ref{fig:sdr_freq}. The $\mathrm{SDR}$s of MM and WMM gradually decreased, but there was no sharp decrease in $\mathrm{SDR}$ appeared in Fig.~\ref{fig:sdr_freq} up to $1500~\mathrm{Hz}$. Therefore, the sharp decrease of the $\mathrm{SDR}$ in Fig.~\ref{fig:sdr_freq} can be considered to be due to the limitation of the estimation accuracy of the expansion coefficients from the pressure measurements at the control points. The $\mathrm{SDR}$s of PM and WPM at $1000~\mathrm{Hz}$ are plotted with respect to the number of control points in Fig.~\ref{fig:sdr_numcp}. In each case, the control points were regularly placed in the target region. To attain $18.4~\mathrm{dB}$ of $\mathrm{SDR}$, 196 control points were necessary for PM although 64 control points were sufficient for WPM owing to the interpolation by the weighting matrix $\bm{W}_{\mathrm{PM}}$. The absolute value of the weighting matrix $|\bm{W}_{\mathrm{PM}}|$for $M=64$ is shown in Fig.~\ref{fig:weight_wpm}. 

MM and WMM does not depend on the control points in this setting. Figure~\ref{fig:sdr_order} shows the $\mathrm{SDR}$ with respect to the maximum order $N_{\mathrm{tr}}$ in the spherical wavefunction expansion. The black line indicates the order of $\lceil kR \rceil$ used as the truncation criterion for MM in the previous experiment (Fig.~\ref{fig:sdr_freq}). From $N_{\mathrm{tr}}=2$ to $14$, the $\mathrm{SDR}$ of MM increased up to around $14.8~\mathrm{dB}$ and it was maintained up to $N_{\mathrm{tr}}=23$. However, above $N_{\mathrm{tr}}=24$, the $\mathrm{SDR}$ of MM sharply decreased. The $\mathrm{SDR}$ of WMM attained $18.4~\mathrm{dB}$ above $N_{\mathrm{tr}}=15$ although it was lower than that of MM between $N_{\mathrm{tr}}=4$ and $12$. Although the excessively large truncation order degenerate the reproduction accuracy in MM, the weighting matrix $\bm{W}_{\mathrm{MM}}$ in WMM appropriately weights on the expansion coefficients to enhance the reproduction accuracy in the target region. The absolute value of the weighting matrix $|\bm{W}_{\mathrm{MM}}|$ at $1000~\mathrm{Hz}$ is shown in Fig.~\ref{fig:weight_wmm}(a) up to $N_{\mathrm{tr}}=7$. The index of $\bm{W}_{\mathrm{MM}}$, denoted by $i$, corresponds to the order $\nu$ and degree $\mu$ as $i=\nu^2+\nu+\mu$. The blue line indicate the range of the indexes of the same $\nu$. The diagonal elements of $|\bm{W}_{\mathrm{MM}}|$ are shown in Fig.~\ref{fig:weight_wmm}(b) by sorting them with respect to $\nu$ and $\mu$. The weights on the expansion coefficients of $\nu=|\mu|$ were relatively larger than those of the other coefficients. Therefore, the empirical weighting scheme of MM, i.e., the extraction of the components of $\nu=|\mu|$, is somehow reasonable. However, the weighting matrix obtained by Eq.~\eqref{eq:weight_wmm} enables achieving much higher reproduction accuracy. 

\subsection{Experiments using real data}

\begin{figure}[t]
\centering
\includegraphics[width=55mm]{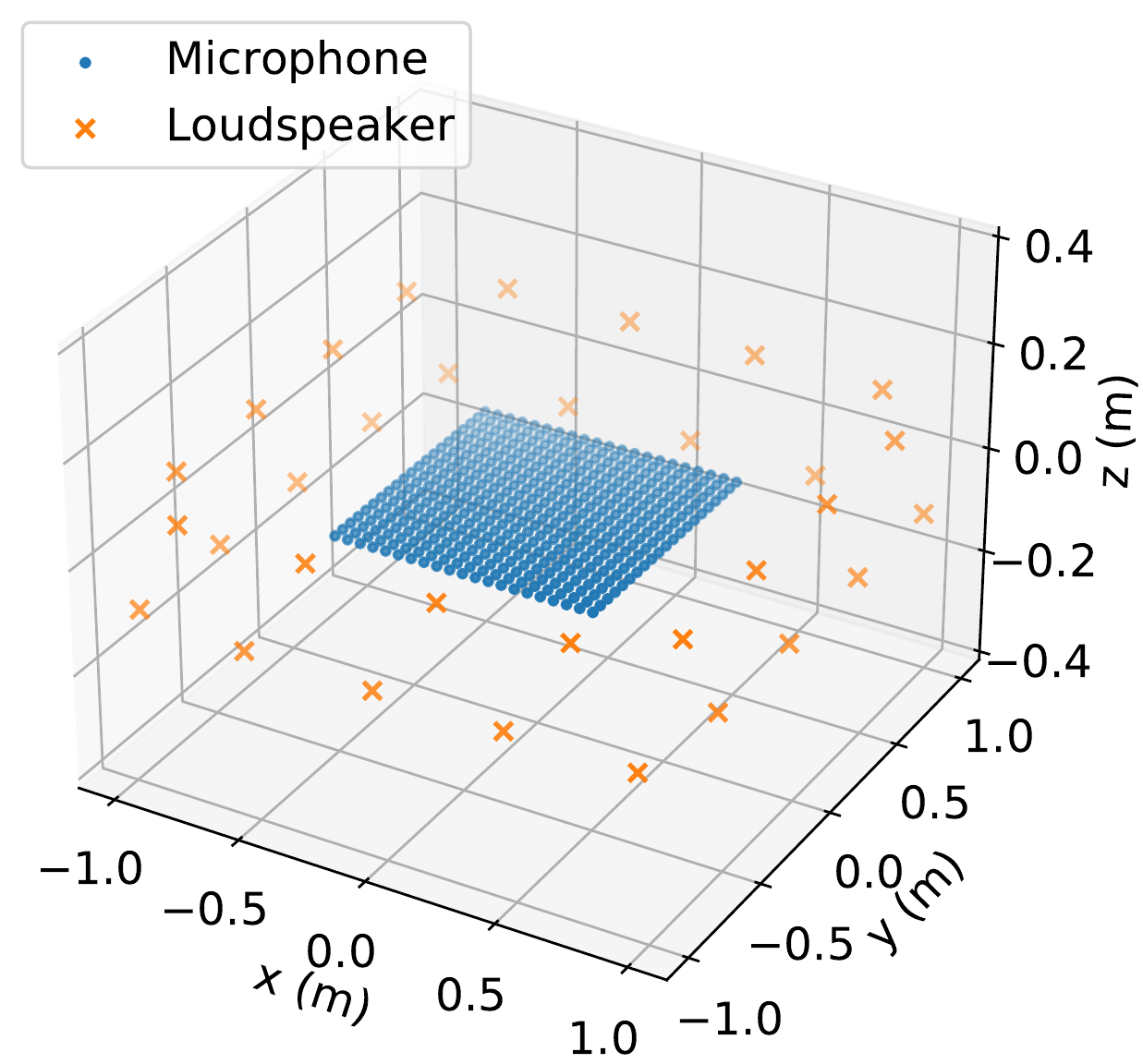}
\caption{Positions of loudspeakers and evaluation points in experiments using real data.}
\label{fig:exp_cond_real}
\end{figure}

\begin{figure}[t]
\centering
\includegraphics[width=65mm]{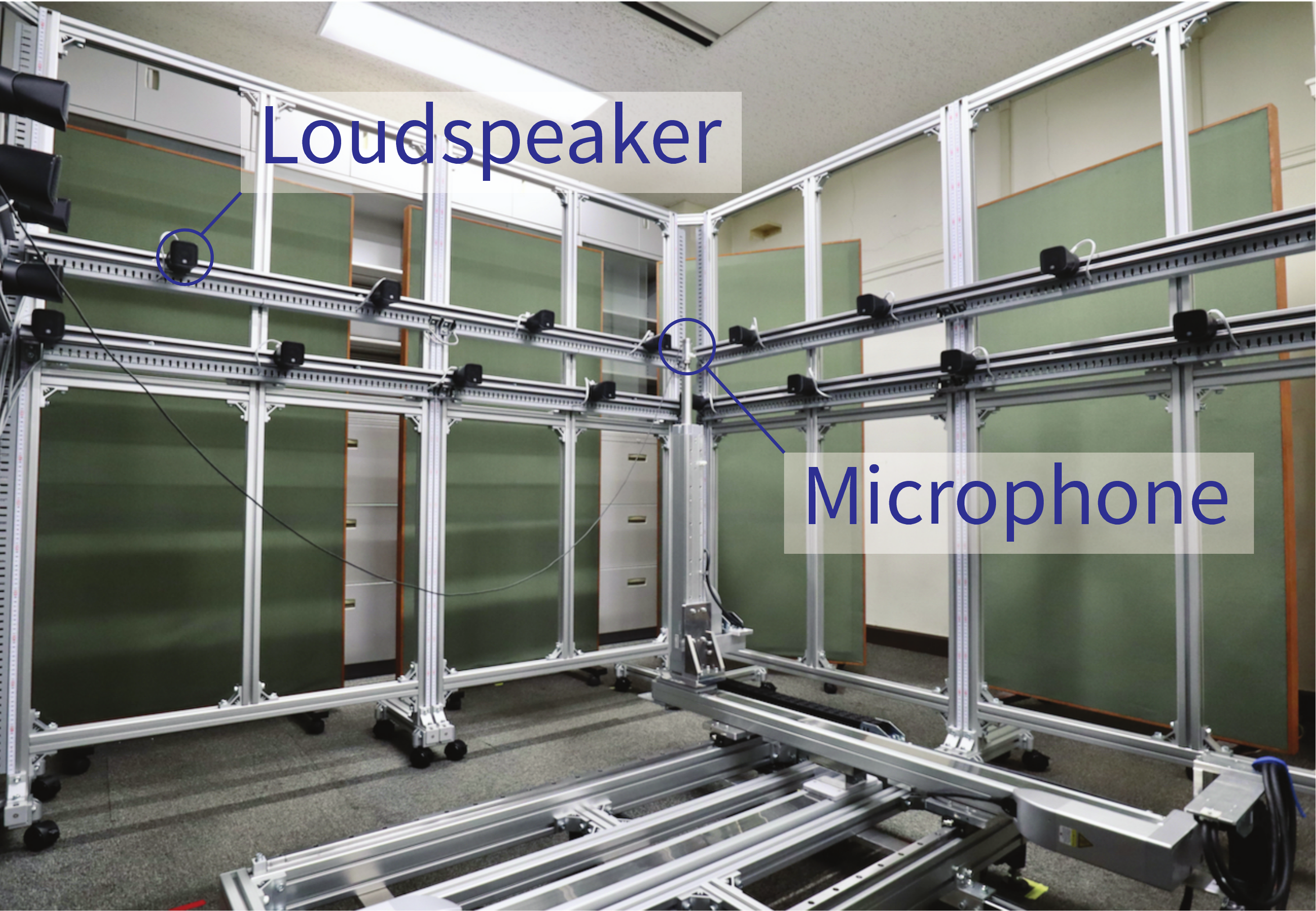}
\caption{Impulse response measurement system.}
\label{fig:imp_meas_system}
\end{figure}

We conducted experiments using impulse responses measured in a practical environment included in the recently published impulse response dataset MeshRIR~\cite{Koyama:WASPAA2021}. The positions of the loudspeakers and evaluation points are shown in Fig.~\ref{fig:exp_cond_real}. Along the borders of two squares with dimensions of $2.0~\mathrm{m}\times2.0~\mathrm{m}$ at heights of $z=-0.2~\mathrm{m}$ and $0.2~\mathrm{m}$, 32 loudspeakers were regularly placed; therefore, 16 loudspeakers were placed along each square. We used ordinary closed loudspeakers (YAMAHA, VXS1MLB). The measurement region was a square with dimensions of $1.0~\mathrm{m} \times 1.0~\mathrm{m}$ at $z=0.0~\mathrm{m}$. The measurement region was discretized at intervals of $0.05~\mathrm{m}$, and $21 \times 21$ ($=441$) evaluation points were obtained; therefore, its spatial Nyquist frequency is around $3400~\mathrm{Hz}$. We measured the impulse response at each evaluation point using an omnidirectional microphone (Primo, EM272J) attached to a Cartesian robot (see Fig.~\ref{fig:imp_meas_system}). The excitation signal of impulse response measurement was a linear swept-sine signal~\cite{Suzuki:JASA1995}. The reverberation time $T_{60}$ was $190~\mathrm{ms}$. The details of the measurement conditions are described in Ref.~\cite{Koyama:WASPAA2021}. The sampling frequency of the impulse responses was $48~\mathrm{kHz}$, but it was downsampled to $8~\mathrm{kHz}$.

We compared the four methods in terms of their reproduction performance in a practical environment. The target region was the same as the region of the evaluation points. Thirty-six microphone positions were regularly chosen from the evaluation points, which were used as control points in PM and WPM and to estimate expansion coefficients of the transfer functions $\mathring{\bm{G}}$ in MM and WMM. The expansion coefficients were estimated up to the 12th order. In MM, the truncation order was set to $N_{\mathrm{tr}}=\min(12, \lceil kR \rceil)$ with $R=0.5\sqrt{2}~\mathrm{m}$ and the expansion coefficients of $\nu=|\mu|$ were only used. Again, the regularization parameter in Eqs.~\eqref{eq:pm_sol}, \eqref{eq:wpm_sol}, \eqref{eq:mm_sol}, and \eqref{eq:wmm_sol} was set as $\sigma_{\max}^2(\bm{A}) \times 10^{-3}$ with the matrix to be inverted $\bm{A}$ at each frequency. The parameter $\xi$ in Eqs.~\eqref{eq:wpm_W_P} and \eqref{eq:est} was set as $\sigma_{\max}(\bm{K})\times 10^{-3}$. We set the desired sound field to a single plane wave propagating to $(\theta,\phi)=(\pi/2,-\pi/4)$. The source signal was a pulse signal whose frequency band was low-pass-filtered up to $900~\mathrm{Hz}$. The filter for obtaining driving signals was designed in the time domain, and its length was $8192~\mathrm{samples}$.
For the evaluation measure in the time domain, we define $\overline{\mathrm{SDR}}$ as
\begin{align}
 \overline{\mathrm{SDR}} = \frac{\iint |u_{\mathrm{des}}(\bm{r},t)|^2 \mathrm{d}\bm{r}\mathrm{d}t}{\iint |u_{\mathrm{syn}}(\bm{r},t) - u_{\mathrm{des}}(\bm{r},t)|^2 \mathrm{d}\bm{r}\mathrm{d}t}.
\end{align}

\begin{figure}[t]
\centering
\subfloat[PM]{\includegraphics[height=30mm,clip]{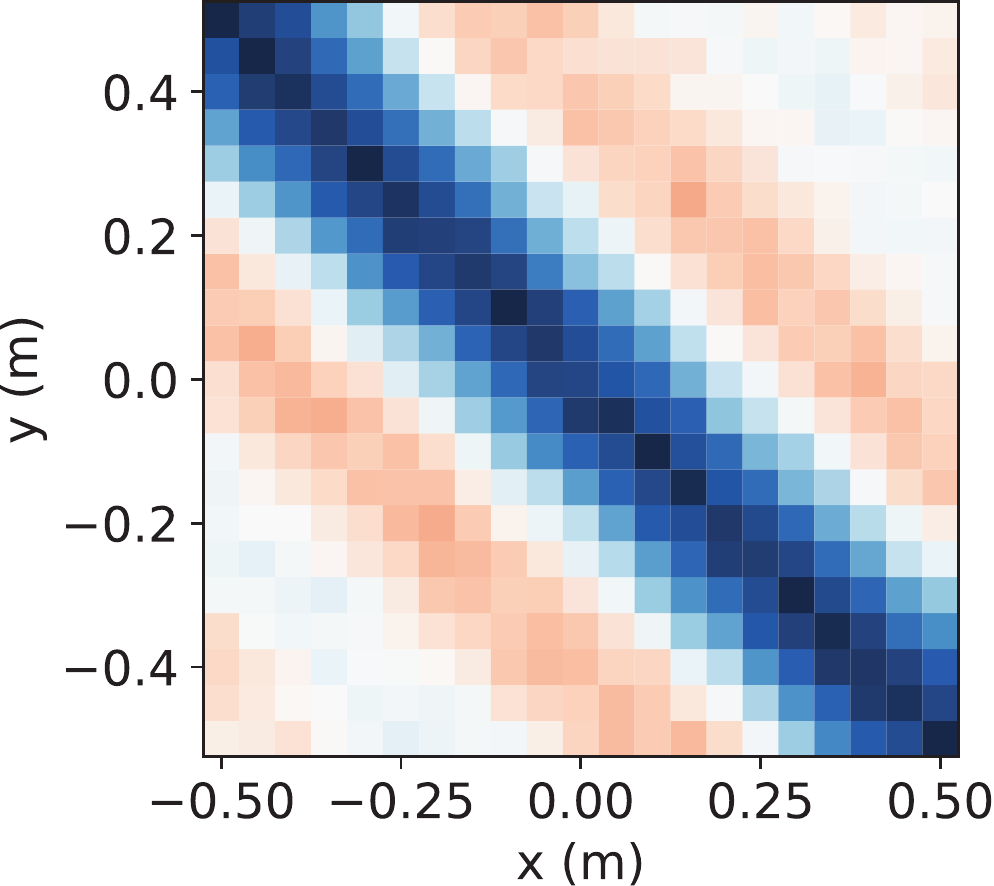} \hspace{10pt}}%
\subfloat[WPM]{\includegraphics[height=30mm,clip]{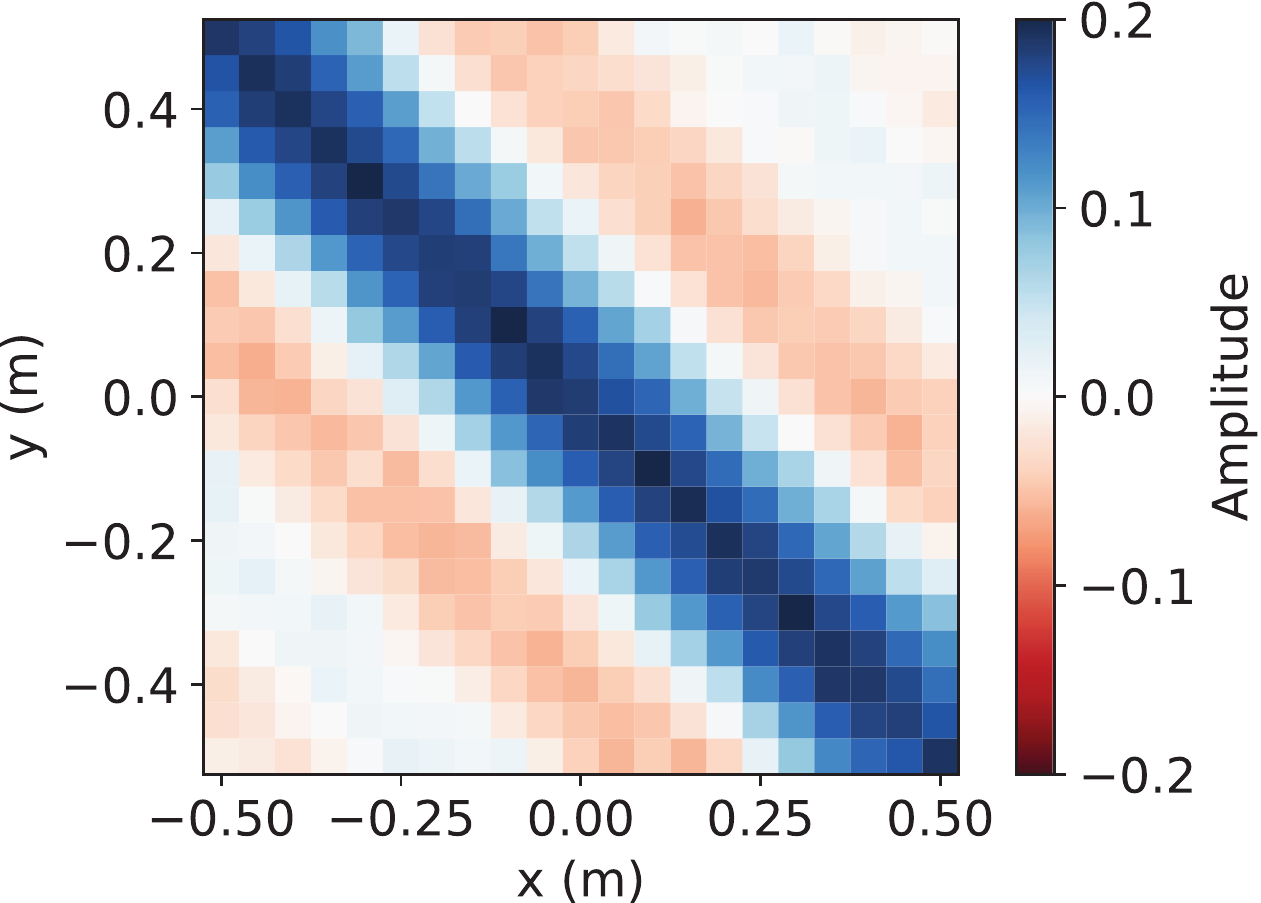}}\\
\subfloat[MM]{\includegraphics[height=30mm,clip]{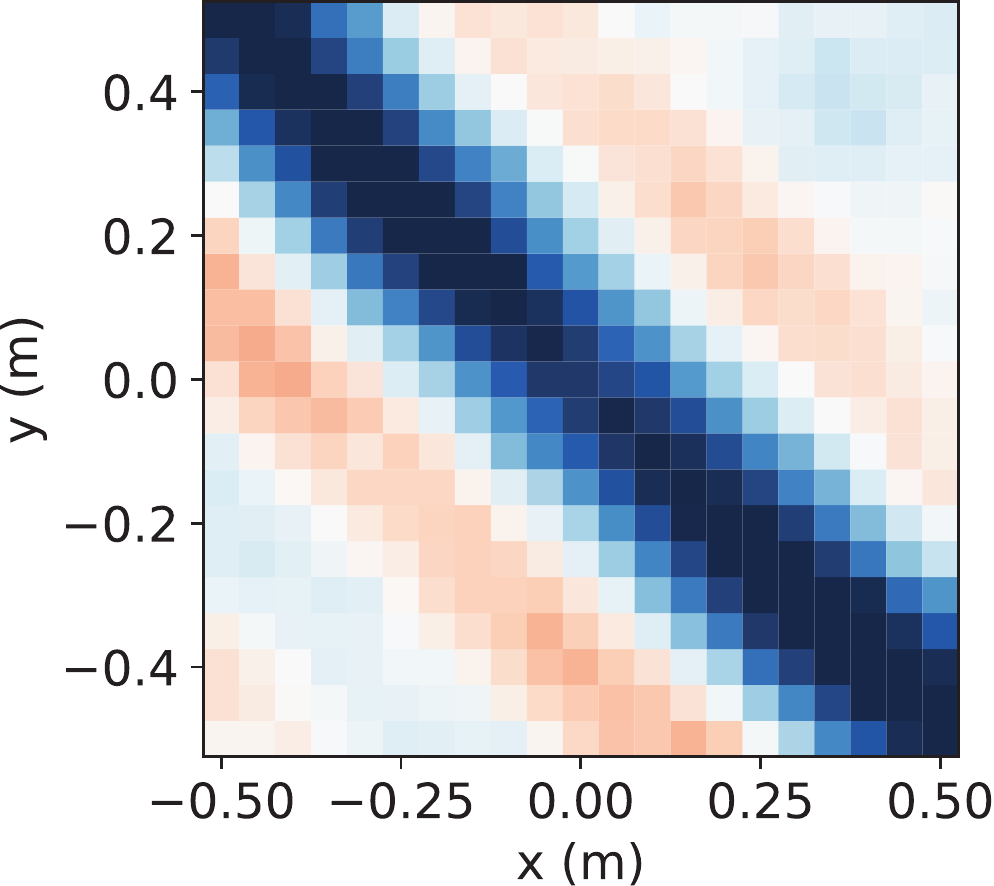} \hspace{10pt}}%
\subfloat[WMM]{\includegraphics[height=30mm,clip]{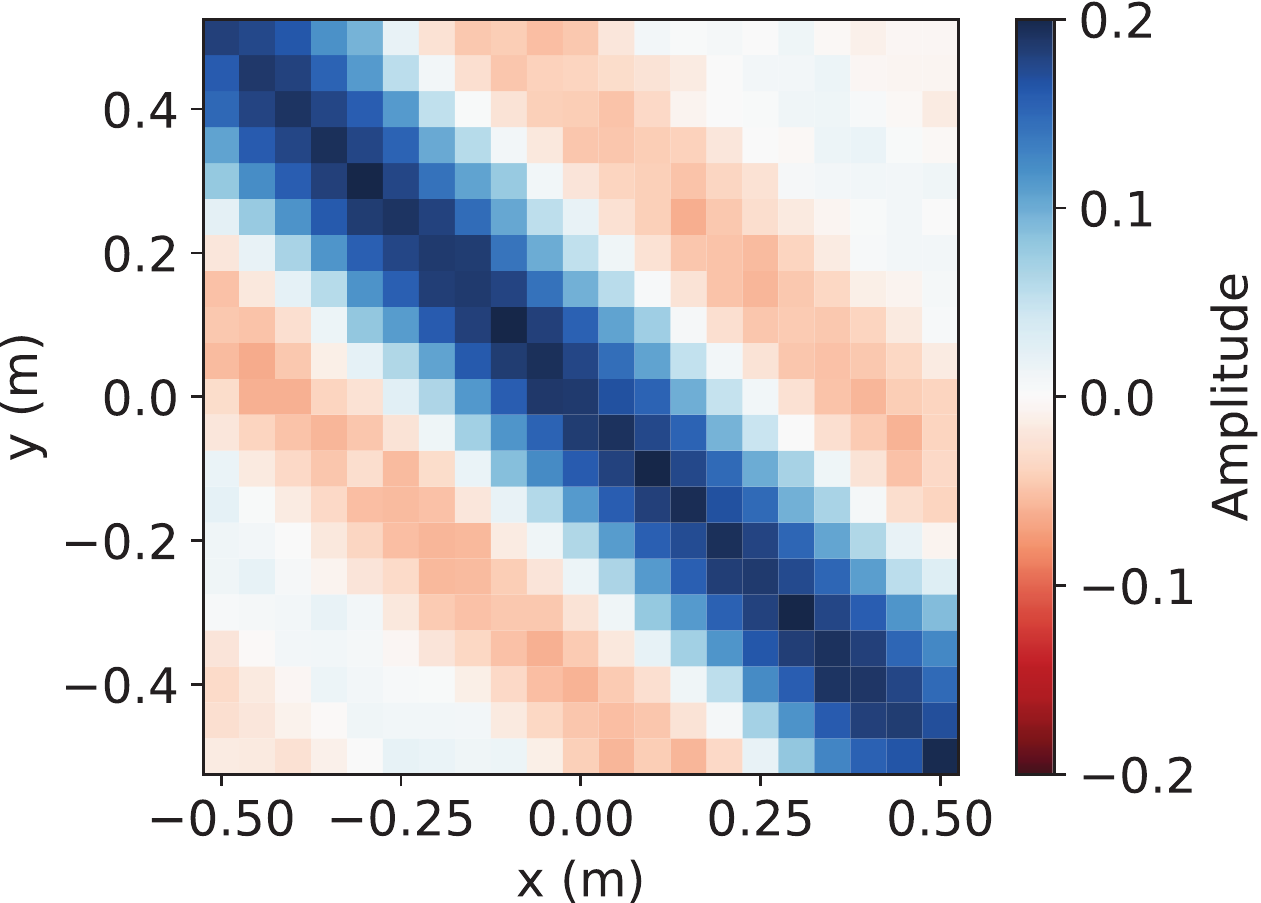}}
\caption{Reproduced pressure distribution at $t=0.51~\mathrm{s}$. $\overline{\mathrm{SDR}}$s of PM, WPM, MM, and WMM were $1.73$, $3.57$, $2.43$, $3.48~\mathrm{dB}$, respectively.}
\label{fig:dist_real}
\end{figure}
\begin{figure}[t]
\centering
\subfloat[PM]{\includegraphics[height=30mm]{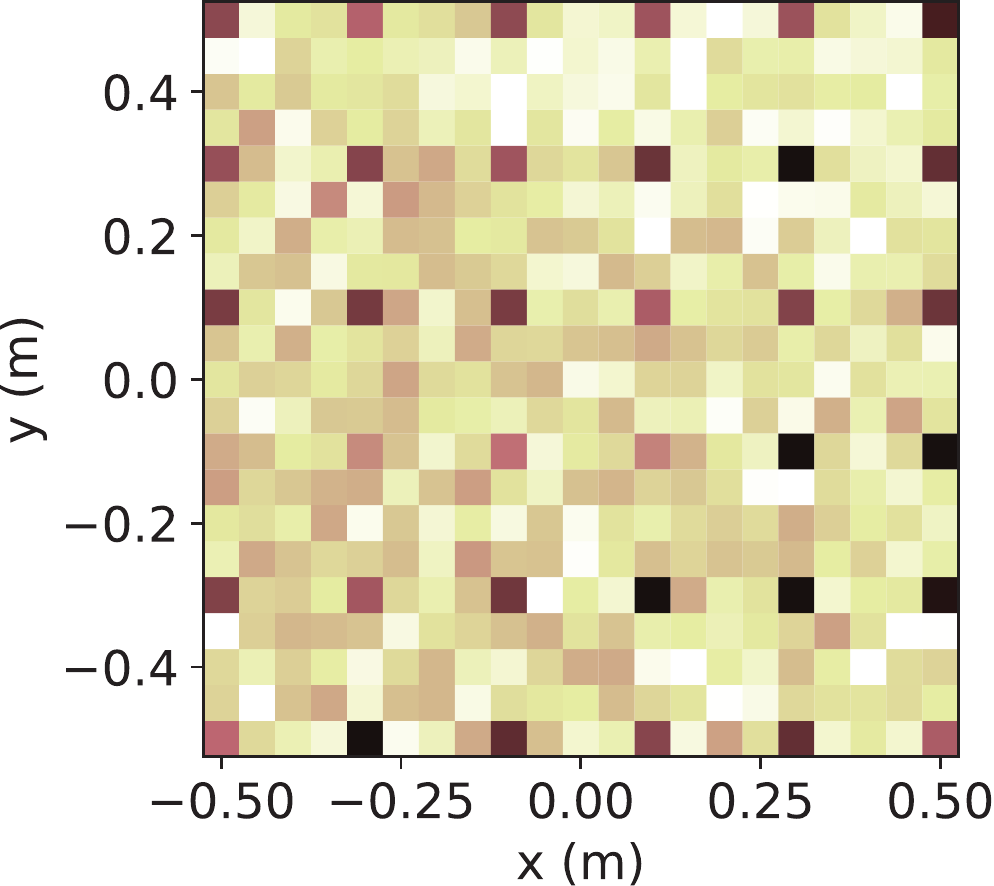} \hspace{10pt}}%
\subfloat[WPM]{\includegraphics[height=30mm]{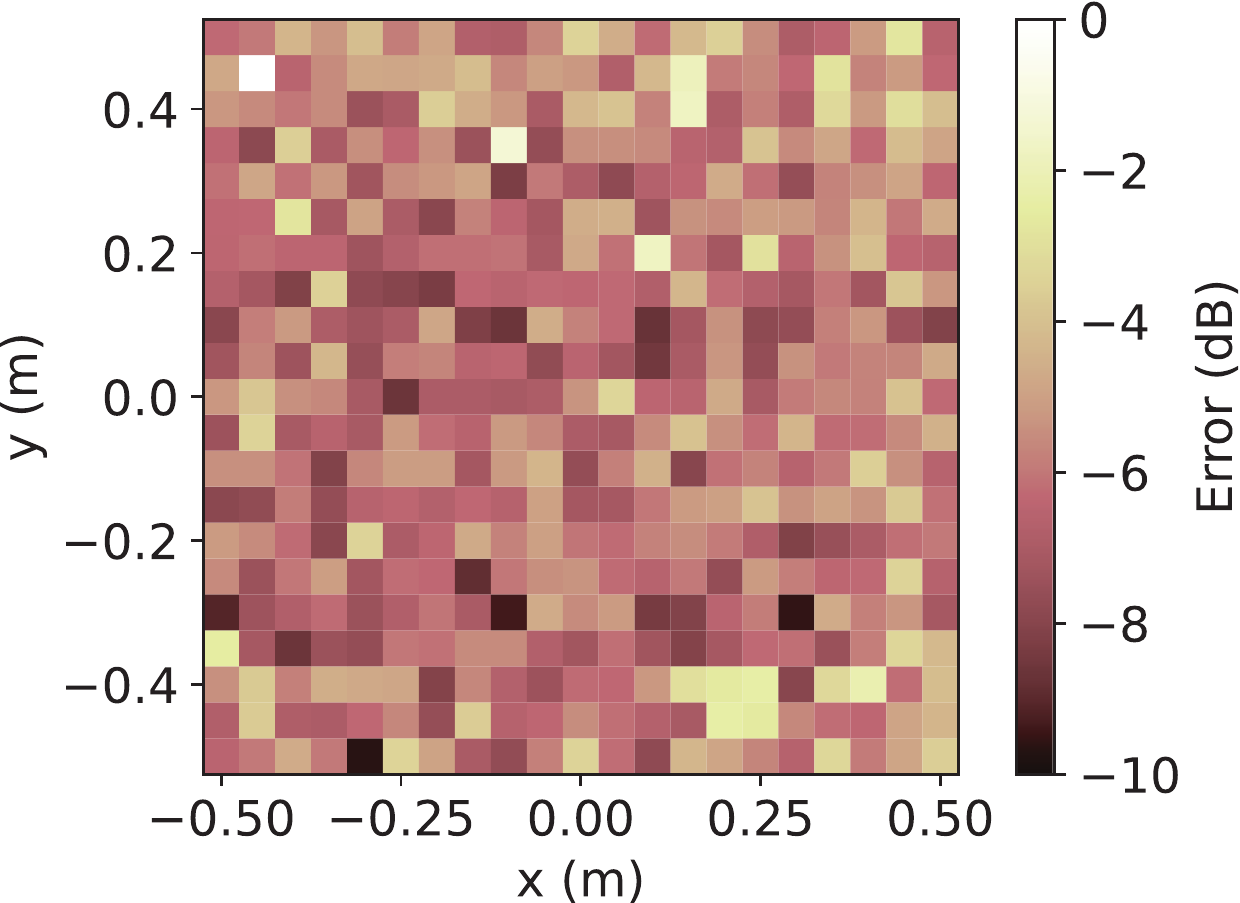}}\\
\subfloat[MM]{\includegraphics[height=30mm]{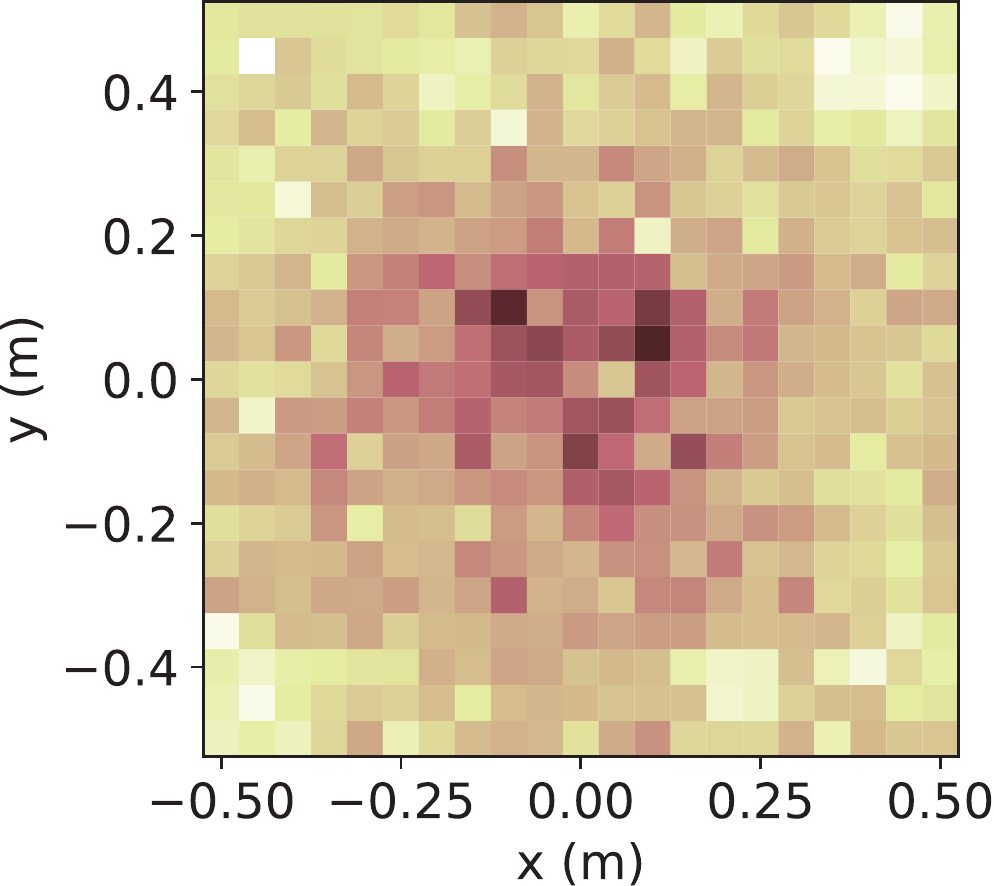} \hspace{10pt}}%
\subfloat[WMM]{\includegraphics[height=30mm]{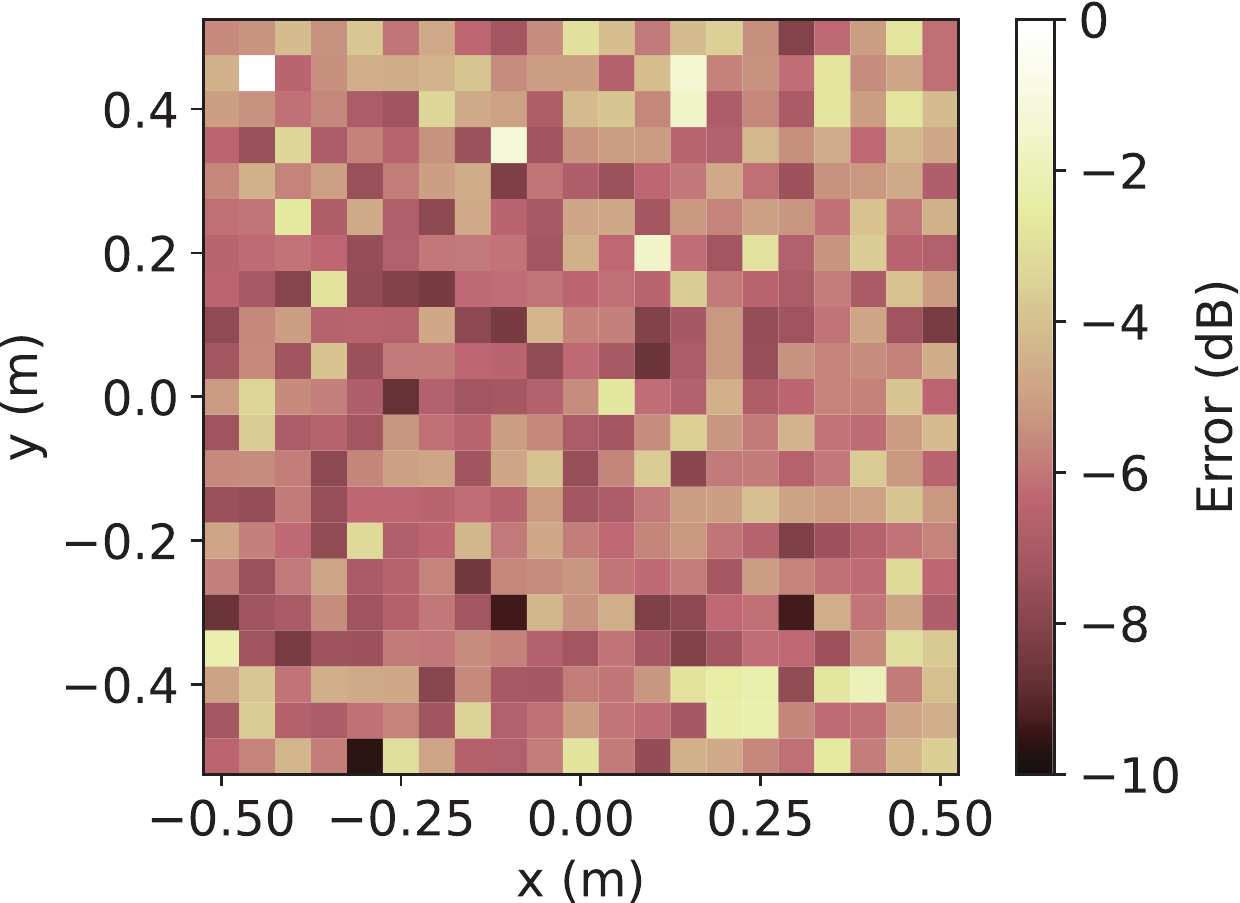}}
\caption{Time-averaged square error distribution.}
\label{fig:err_real}
\end{figure}

Figure~\ref{fig:dist_real} shows the reproduced pressure distributions at $t=0.51~\mathrm{s}$. Time-averaged square error distributions are shown in Fig.~\ref{fig:err_real}. In PM, a small time-averaged square error was observed at the positions of the control points, but the region between them contains large errors. The time-averaged square error of MM was small around the center of the target region, but that was high in the off-center region. In WPM and WMM, a small square error was obtained over the target region. $\overline{\mathrm{SDR}}$s of PM, WPM, MM, and WMM were $1.73$, $3.57$, $2.43$, $3.48~\mathrm{dB}$, respectively.

\section{DISCUSSION}
\label{sec:discuss}

The weighting matrices in the weighted pressure and mode matching, $\bm{W}_{\mathrm{PM}}$ and $\bm{W}_{\mathrm{MM}}$, were derived to enhance the reproduction accuracy of pressure and mode matching. Although the simple formulations were only shown to discuss the relationship between the two methods, the reproduction accuracy can be further enhanced by introducing directional weighting for sound field capturing and/or regional weighting for sound field reproduction~\cite{Ueno:IEEE_ACM_J_ASLP2019,Ueno:IEEE_J_SP_2021,Koyama:ICA2022}. We here discuss the difference between the two methods in detail.

Although the cost functions of the weighted pressure and mode matching are similar, the roles of the weighing matrices are different. The weighting matrix $\bm{W}_{\mathrm{PM}}$ in the weighted pressure matching is derived from the interpolation of the pressure field between the control points based on the kernel ridge regression to alleviate the effect of spatial aliasing artifacts owing to the spatial sampling in the target region. In contrast, the weighted mode matching is formulated based on the spherical wavefunction expansion with the given expansion coefficients of the transfer functions and desired field. Therefore, the weighted mode matching, as well as mode matching, does not suffer from spatial aliasing owing to the sound field capturing as long as the accurate expansion coefficients are given. The weighting matrix $\bm{W}_{\mathrm{MM}}$ is derived from the approximation of the original cost function $J$ in Eq.~\eqref{eq:cost} instead of simply matching the expansion coefficients up to an empirical truncation order.

However, in practical situations, the expansion coefficients of the transfer functions $\mathring{\bm{G}}$ must be estimated from the microphone measurements because it is difficult to accurately model the practical loudspeakers and reverberations without the measurements. The expansion coefficients of the desired field $\mathring{\bm{u}}^{\mathrm{des}}$ must also be estimated from the discrete set of measurements when their analytical representations are difficult to obtain. The infinite-dimensional harmonic analysis is one of the methods to estimate the expansion coefficients from the measurements. As shown in Section~\ref{sec:rel}, when the expansion coefficients $\mathring{\bm{G}}$ and $\mathring{\bm{u}}^{\mathrm{des}}$ in the weighted mode matching are estimated from the pressure observations at the control points by the infinite-dimensional harmonic analysis, the weighted mode matching corresponds to the weighted pressure matching. In the experiments, the reproduction accuracy of these two methods were almost identical. Since the computation of $\bm{W}_{\mathrm{PM}}$ is generally simpler than that of $\bm{W}_{\mathrm{MM}}$ and the estimation operator of the infinite-dimensional harmonic analysis in Eq.~\eqref{eq:est}, the weighted pressure matching is simpler for implementation compared to the weighted mode matching. However, the weighted pressure matching is applicable only when the pressure measurements at the control points are available because the kernel function is derived for interpolating the pressures. When the microphones have directivity, the infinite-dimensional harmonic analysis can be applied.

Another difference is the number of parameters to represent the sound field. It has been shown that the number of expansion coefficients required for the weighted mode matching can be smaller than the number of control points required for pressure matching when the target region is sphere in Ref.~\cite{Ueno:IEEE_ACM_J_ASLP2019} (see Fig.~4 in Ref.~\cite{Ueno:IEEE_ACM_J_ASLP2019}). When the target region is not a sphere, for example, a horizontal plane, as in the experiments, the representation by the spherical wavefunction expansion is sometimes redundant, and that is the reason why mode matching does not perform well in the experiments. In the experiment in Section~\ref{sec:exp_sim}, the maximum order $N_{\mathrm{tr}}=15$ required to attain $18.4~\mathrm{dB}$ of $\mathrm{SDR}$ in the weighted mode matching corresponds to 256 expansion coefficients, which is much larger than the number of control points, 64, required to attain the same $\mathrm{SDR}$ in the weighted pressure matching. The number of control points can be further reduced by the sensor placement methods~\cite{Koyama:IEEE_ACM_J_ASLP2020}. However, the weighting matrix $\bm{W}_{\mathrm{MM}}$ of the weighted mode matching is significantly sparse, as shown in Fig.~\ref{fig:weight_wmm}. By extracting the columns and rows of the index set $\{ k \ | \ \sum_i \sum_k |\bm{W}_{\mathrm{MM},i,k}| +  \sum_j  \sum_k |\bm{W}_{\mathrm{MM},k,j}| > \delta \}$ with $\delta=\max(|\bm{W}_{\mathrm{MM}}|) \times 10^{-3}$, the number of expansion coefficients was reduced to 120 with the same $\mathrm{SDR}$. Therefore, it is possible to extract required expansion coefficients based on the weighting matrix $\bm{W}_{\mathrm{MM}}$ to reduce the number of parameters to represent the sound field. In addition, the expansion coefficients of the spherical wavefunctions are compatible with the existing ambisonics format. Their independence from the microphone positions as an intermediate representation is useful for storing and transmitting data.

Although we focused on the relationship between the weighted pressure and mode matching, a common issue for the sound field reproduction methods including both the analytical and numerical methods is spatial aliasing owing to the discrete arrangement of the secondary sources. Although this issue is beyond the scope of this paper, we here briefly discuss the spatial aliasing problem in the sound field reproduction. Based on the single layer potential~\cite{Colton:InvAcoust_2013}, any source-free sound field in the interior target region can be synthesized by continuously distributed point sources on a surface surrounding the target region. Since the continuous distribution is replaced with a discrete set of secondary sources in practice, the reproduction accuracy can deteriorate at high frequencies. Specifically, degradation in sound localization and coloration of reproduced sounds can occur. In some applications such as local-field reproduction and noise cancellation, the reproduced frequency range is targeted at low frequencies; therefore, the required number of secondary sources for accurate reproduction is relatively small. The sound field reproduction for the audible frequency range requires a large number of secondary sources. 
Several attempts have been made to combine with other spatial audio reproduction techniques for high frequencies~\cite{Kamado:ICASSP2011} to prioritize the flat amplitude response under the assumption that inaccurate phase distribution is acceptable at high frequencies in the human auditory system.
Nevertheless, there are several techniques to further reduce the number of secondary sources. The first technique is to reduce the number of parameters to be controlled, which makes the problem to be solved in the (weighted) pressure and mode matching overdetermined even with the small number of secondary sources. For example, by limiting the range of the target region and introducing the regional importance of reproduction, the number of control points or expansion coefficients to be controlled can be reduced. As in the experiments, the target region is frequently limited to the horizontal plane because the listeners' ears can be assumed to be approximately on the same plane in practical situations. The second technique is the optimization of the secondary source placement~\cite{Khalilian:IEEE_ACM_J_ASLP_2016,Koyama:IEEE_ACM_J_ASLP2020,Kimura:WASPAA2021}. By selecting an optimal set of secondary source positions from candidate positions in a certain criterion, the minimum required number of secondary sources and their optimal placement can be obtained. We consider that spatial aliasing owing to the secondary sources is still an open issue in this field.

\section{CONCLUSION}
\label{sec:conclusion}

Theoretical and experimental comparisons of two sound field reproduction methods, weighted pressure and mode matching, were carried out, which can be regarded as a generalization of conventional pressure and mode matching, respectively. In the weighted pressure matching, the weighting matrix is obtained on the basis of the kernel interpolation of the sound field from the pressure at the control points. The weighted mode matching is derived on the basis of the spherical wavefunction expansion of the sound field, and the weighting matrix is defined as the regional integration of the spherical wavefunctions. When the expansion coefficients of the desired sound field and transfer functions are estimated from the pressure observations at the control points by infinite-dimensional harmonic analysis, the weighted mode matching corresponds to the weighted pressure matching. In this sense, the weighted mode matching is more general than the weighted pressure matching because the desired sound field can be given as the analytical formulation of expansion coefficients and directional microphones can also be used to estimate the expansion coefficients. The advantage of the weighted pressure matching is its simplicity for implementation. 
The difference in the number of parameters required to represent the sound field is discussed through the experiments. The redundancy of the spherical wavefunction expansion when the target region is not a sphere can be alleviated to some extent by extracting the expansion coefficients based on the weighting matrix of the weighted mode matching.

\vfill

\section{ACKNOWLEDGMENT}
This work was supported by JST FOREST Program, Grant Number JPMJFR216M, and  JSPS KAKENHI, Grant Number 22H03608.


\bibliography{str_def_abrv,koyama_en,refs}
\bibliographystyle{jaes.bst}

\break

\appendix


\section*{REPRESENTATION OF OBSERVED SIGNAL}

The detailed derivation of Eq.~\eqref{eq:sm_ip} is described, which is also shown in Refs.~\cite{Ueno:IEEE_SPL2018,Ueno:IEEE_J_SP_2021}. First, the sound field $u(\bm{r})$ in Eq.~\eqref{eq:sphwave_exp} can also be represented by plane wave expansion around the expansion center $\bm{r}_{\mathrm{o}}$ as
\begin{align}
 u(\bm{r}) = \int_{\bm{\eta} \in \mathbb{S}_2} \tilde{u} (\bm{\eta}; \bm{r}_{\mathrm{o}}) 
 \mathrm{e}^{-\mathrm{j} \bm{\eta}^{\mathsf{T}} (\bm{r}-\bm{r}_{\mathrm{o}})} \mathrm{d}\bm{\eta},
\end{align}
where $\tilde{u}$ is the planewave weight and $\bm{\eta}$ denotes the arrival direction of the plane wave defined on the unit sphere $\mathbb{S}_2$. Suppose that a microphone with the directivity pattern $c(\bm{\eta})$ is placed at $\bm{r}_{\mathrm{o}}$. Then, the microphone's response $s$ is given by
\begin{align}
 s = \int_{\bm{\eta} \in \mathbb{S}_2} \tilde{u}(\bm{\eta};\bm{r}_{\mathrm{o}}) c(\bm{\eta}) \mathrm{d} \bm{\eta}.
\end{align}
When the spherical wavefunction $\varphi_{\nu,\mu}(\bm{r}-\bm{r}_{\mathrm{o}})$ is represented by plane wave expansion, its weight $\tilde{\varphi}_{\nu,\mu}(\bm{\eta}; \bm{r}_{\mathrm{o}}) $ is obtained from the Funk--Hecke formula~\cite{Martin:MultScat} as 
\begin{align}
 \tilde{\varphi}_{\nu,\mu}(\bm{\eta}; \bm{r}_{\mathrm{o}}) = \frac{\mathrm{j}^{\nu}}{\sqrt{4\pi}} Y_{\nu,\mu}(\bm{\eta}).
\end{align}
Therefore, the microphone's response of the sound field represented by Eq.~\eqref{eq:sphwave_exp} is described as
\begin{align}
 s &= \sum_{\nu=0}^{\infty} \sum_{\mu=-\nu}^{\nu} \mathring{u}_{\nu,\mu}(\bm{r}_{\mathrm{o}}) \int_{\bm{\eta}\in\mathbb{S}_2} \frac{\mathrm{j}^{\nu}}{\sqrt{4\pi}} Y_{\nu,\mu}(\bm{\eta}) c(\bm{\eta}) \mathrm{d} \bm{\eta} \notag\\
&= \sum_{\nu=0}^{\infty} \sum_{\mu=-\nu}^{\nu} \mathring{u}_{\nu,\mu}(\bm{r}_{\mathrm{o}}) c_{\nu,\mu}^{\ast},
\end{align}
where 
\begin{align}
 c_{\nu,\mu} &= \frac{(-\mathrm{j})^{\nu}}{\sqrt{4\pi}} \int_{\bm{\eta}\in\mathbb{S}_2} c(\bm{\eta})^{\ast} Y_{\nu,\mu}(\bm{\eta})^{\ast} \mathrm{d}\bm{\eta}. 
\end{align}
Therefore, the observed signal is represented by Eq.~\eqref{eq:sm_ip}. 
Note that generally-used directivity patterns are represented by low-order coefficients of $c_{\nu,\mu}$. For example, the directivity pattern of unidirectional microphone is represented as
\begin{align}
 c(\bm{\eta},\bm{\eta}_{\mathrm{dir}}) = \beta + (1-\beta) \bm{\eta} \cdot \bm{\eta}_{\mathrm{dir}} 
\end{align}
with the constant $\beta \in [0,1]$ and direction of the microphone (peak of directivity) $\bm{\eta}_{\mathrm{dir}}$.
Hence, $c_{\nu,\mu}$ is obtained as
\begin{align}
 c_{\nu,\mu} = 
 \begin{cases}
  \beta & \nu=0, \mu=0 \\
  \frac{\sqrt{4\pi} \mathrm{j}}{3} (1-\beta) Y_{-1}^1(\bm{\eta}_{\mathrm{dir}})^{\ast} & \nu=1, \mu=-1 \\
  \frac{\sqrt{4\pi} \mathrm{j}}{3} (1-\beta) Y_0^1(\bm{\eta}_{\mathrm{dir}})^{\ast} & \nu=1, \mu=0 \\
  \frac{\sqrt{4\pi} \mathrm{j}}{3} (1-\beta) Y_1^1(\bm{\eta}_{\mathrm{dir}})^{\ast} & \nu=1, \mu=1 \\
  0 & \text{otherwise}
 \end{cases},
\end{align}
which corresponds to the case of the omnidirectional microphone Eq.~\eqref{eq:c_omni} by setting $\beta = 1$.

%
%
\end{document}